\documentclass[article,11pt]{emulateapj}
\usepackage{graphicx}
\usepackage{natbib}
\usepackage{amsmath}
\usepackage{amssymb}
\usepackage{graphicx}
\def\jgr{J. Geophys. Res. }

\def\ao{Appl. Opt. }
\def\aj{Astron. J.}
\def\apj{Astrophys. J.}
\def\grl{Geophys. Rev. Let.}
\def\jqsrt{J. Quant. Spectrosc. \& Rad. Transf. }

\def\planss{Planet. \& Space Sci.}
\def\hto{H$_2$O }

\def\icm{cm$^{-1}$ }

\def\deg{$^\circ$ }

\def\mum{$\mu$m }
\def\mumx{$\mu$m}
\def\chf{CH$_4$ }
\def\chfx{CH$_4$}
\def\chtd{CH$_3$D }

\def\chisq{$\chi^2$ }

\def\pht{PH$_3$ }
\def\phtx{PH$_3$}

\def\pthfx{P$_2$H$_4$}
\def\nht{NH$_3$ }
\def\nhtx{NH$_3$}
\def\nhfsh{NH$_4$SH }
\def\nhfshx{NH$_4$SH}
\def\nthf{N$_2$H$_4$ }
\def\nthfx{N$_2$H$_4$}

\begin{document}
\title{Saturn's Great Storm of 2010-2011:
    Evidence for ammonia \\ and water ices from analysis of
    VIMS spectra.}
\author{L.A. Sromovsky, K.~H. Baines, and P.~M. Fry} 
\affil{Space Science and Engineering Center, University of Wisconsin-Madison\\
1225 West Dayton Street, Madison, WI 53706}
\slugcomment{Journal reference: Icarus 226 (2013) 402-418.}

\begin{abstract}

Our analysis of Cassini/VIMS near-infrared spectra of Saturn's Great Storm
of 2010-2011 reveals a multi-component aerosol composition comprised
primarily of ammonia ice, with a significant component of water
ice. The most likely third component is ammonium hydrosulfide or some
weakly absorbing material similar to what dominates visible clouds
outside the storm region.  Horizontally heterogeneous models favor
ammonium hydrosulfide as the third component, while horizontally
uniform models favor the weak absorber.  Both models rely on water ice
absorption to compensate for residual spectral gradients produced by
ammonia ice from 3.0 \mum to 3.1 \mum and need the third component to
fill in the sharp ammonia ice absorption peak near 2.96 \mumx. The
best heterogeneous model has spatial coverage fractions of 55\%
ammonia ice, 22\% water ice, and 23\% ammonium hydrosulfide. The best
homogeneous model has an optically thin layer of weakly absorbing
particles above an optically thick layer of water ice particles coated
by ammonia ice.  This is the first spectroscopic evidence of water ice
in Saturn's atmosphere, found near the level of Saturn's visible cloud
deck where it could only be delivered by powerful convection
originating from $\sim$200 km deeper in the atmosphere.

\end{abstract}

\maketitle
\shortauthors{Sromovsky et al.} 
\shorttitle{Composition of Saturn's storm cloud.}
\newpage

\section{Introduction}
\begin{figure*}[!htb]\centering
\includegraphics[width=5.in]{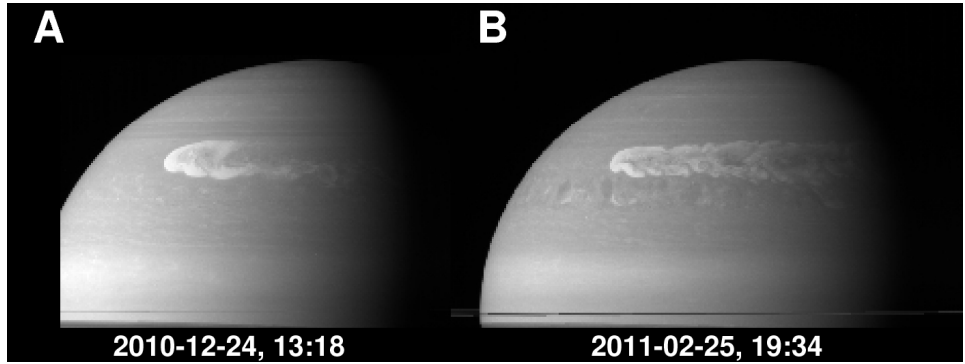}
\caption{ Cassini ISS images of Saturn's Great Storm of
2010-2011 ({\bf A}) at an early stage of development in December 2010
and ({\bf B}) at approximately the same time as the first VIMS
spectral images obtained in February 2011. Both images were made with
a CB2 filter (752 nm) and the wide angle camera.}
\label{Fig:ISSimages}
\end{figure*}

\begin{figure*}[!htb]\centering
\includegraphics[width=3.1in]{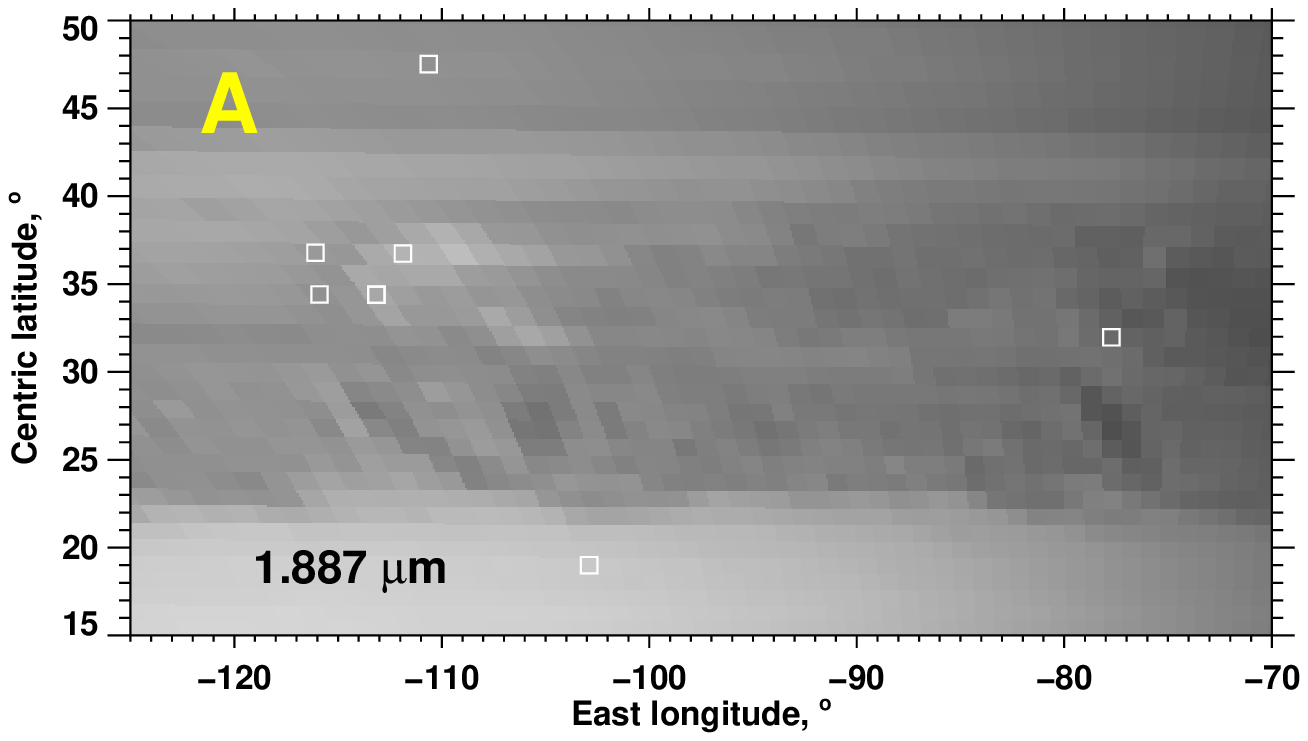}
\includegraphics[width=3.1in]{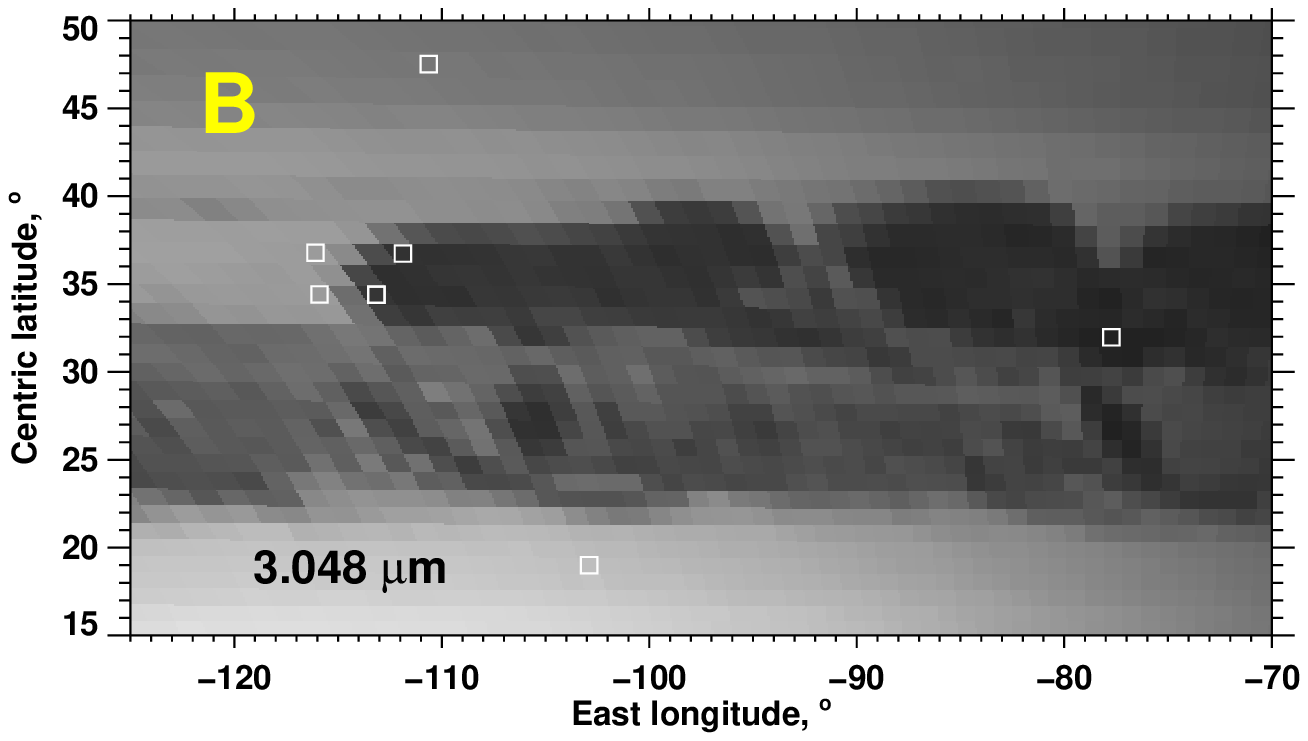}\\
\vspace{-.04in}\includegraphics[width=3.1in]{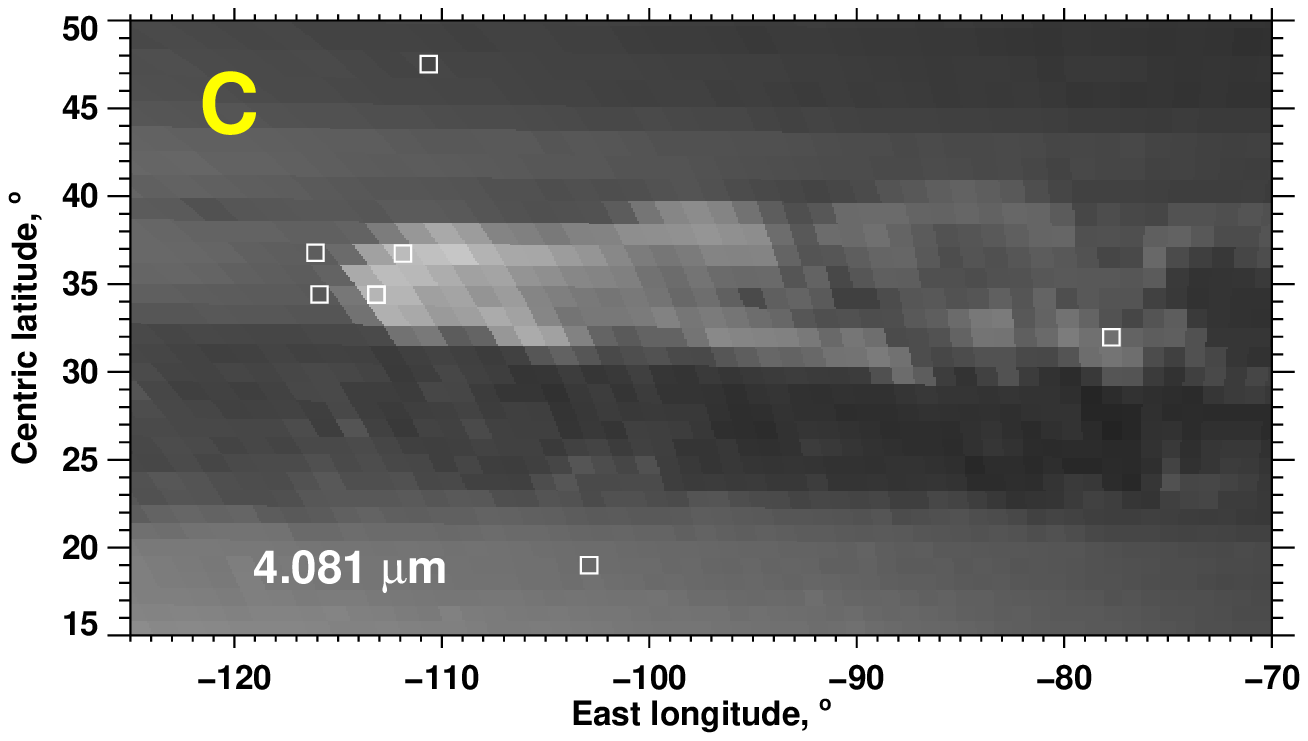}
\includegraphics[width=3.1in]{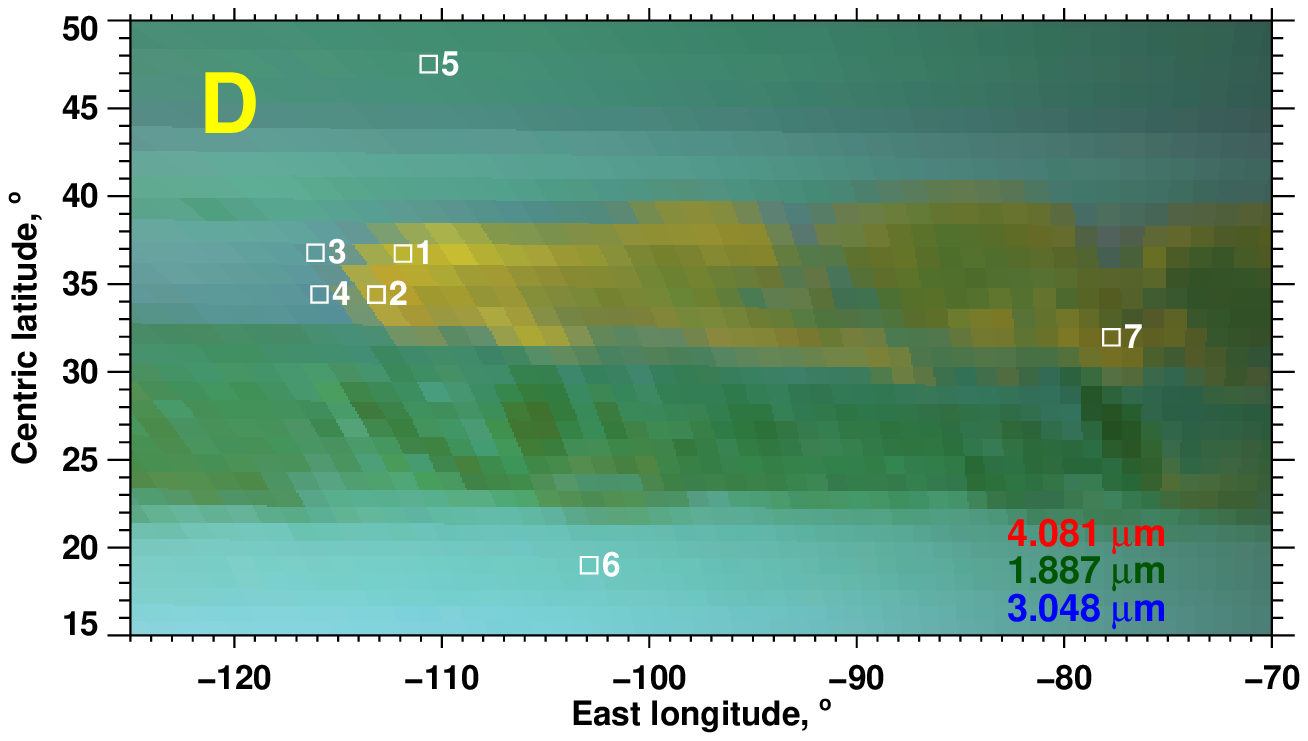}\\
\vspace{-.1in}\includegraphics[width=6.2in]{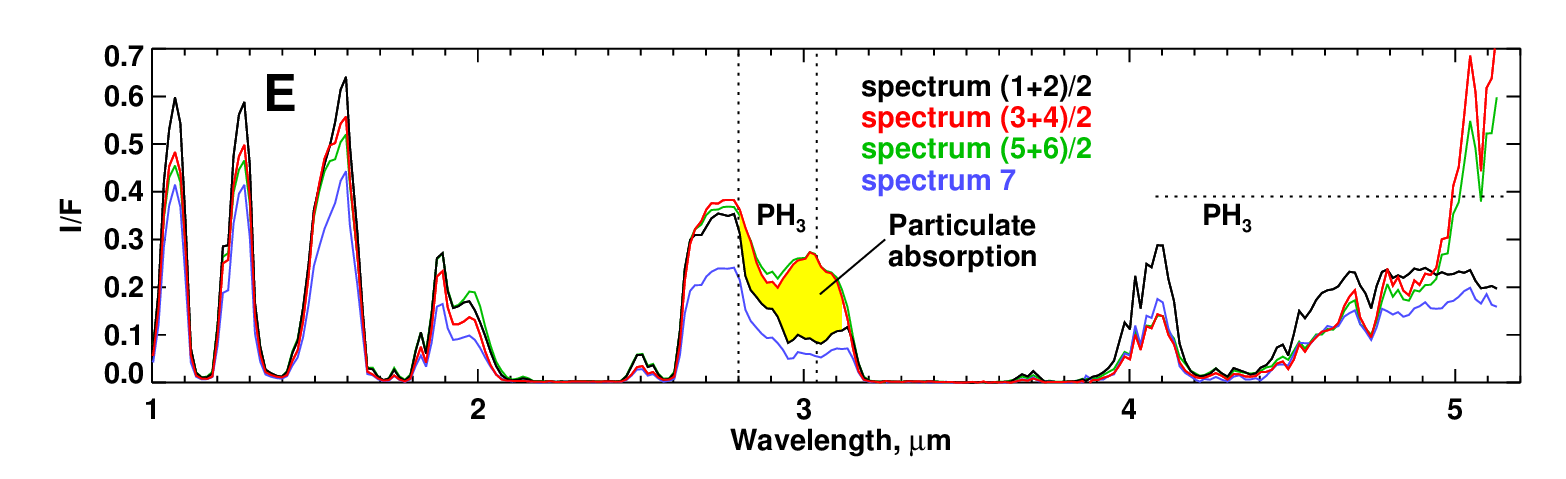}\\
\caption{({\bf A-C}) VIMS spectral images of the storm in February
  2011, sampled at wavelengths indicated within each panel, remapped
  to an orthogonal projection, and linearly stretched so that white
  corresponds to a reflectivity (I/F) of 0.4. The color composite
  ({\bf D}) is created from spectral images at 3.048 \mum (blue
  component), 1.887 \mum (green component), and 4.081 \mum (red
  component).  Numbered boxes indicate the locations at which spectral
  samples ({\bf E}) were obtained. Here closely similar spectra are
  averaged to simplify the figure.  Strong absorption at 3.048 \mum
  reduces the blue contribution in the color composite image, making
  the storm feature appear orange. The parts of the storm that are
  bright at 4.081 \mum (locations 1 and 2, for example) are much
  darker at 3.048 \mum than the region ahead of the storm (locations 3
  and 4).  1.887 \mum and 4.081 \mum are pseudo-continuum wavelengths,
  subject to little gas absorption, and thus show reflective
  properties of clouds more directly. The strong reflectivity at 4.081
  \mum indicates that the cloud particles are relatively large
  ($\sim$1 \mum or more). The spectra from regions 3-6 are very
  similar, especially in the 2.9-3.04 \mum region where they contain a
  similar phosphine absorption feature. The difference between those
  spectra and the storm head spectra (1 and 2) is colored yellow to
  show the extra absorption that is mainly due to particulates.}
\label{Fig:VIMSimages}
\end{figure*}

Beginning in early December of 2010 amateur astronomers around the
world observed the outbreak of a remarkably large storm in the
atmosphere of Saturn \citep{Sanchez-Lavega2011Nature}.  This large
feature is the sixth known of a class of features historically
referred to as Great White Spots, which have occurred at roughly
thirty year intervals \citep{Sanchez-Lavega1991}, i.e. at intervals
close to one Saturnian year, always in the northern hemisphere, and
usually in that hemisphere's summer.
This is the first such storm observed close up by a spacecraft
(Cassini) in orbit around Saturn. The Cassini Imaging Science
Subsystem (ISS) captured images of the storm beginning in December
(Fig.\ \ref{Fig:ISSimages}) and the Visual and Infrared Mapping
Spectrometer (VIMS) captured its first spectra of the storm in
February 2011 (Fig.\ \ref{Fig:VIMSimages}).
It was then centered at 35\deg N and covered about 7\deg in latitude,
with a long and widening downstream plume.
The vertical and horizontal scale of its effects were probed
by thermal infrared spectroscopy, which showed that within a month the storm
generated large perturbations of atmospheric temperatures, winds, and
composition over a wide area between 20\deg N and 50\deg N
\citep{Fletcher2011Sci}.
 Dynamical models of such storms
\citep{Hueso2004Icar} suggest that they originate at the 10-12 bar
level where water vapor condensation is expected, and generate
enormous convective towers reaching pressures as low as $\sim$150 mb.
It is thus conceivable that these storms provide a window into the
deeper layers of Saturn's atmosphere.

The near-IR spectral observations of VIMS (from 1 \mum to 5.15 \mumx)
are of special interest because they are particularly diagnostic of
cloud particle composition.
It was immediately obvious from VIMS spectra that the cloud particles
in this storm had an unusual composition; they displayed a
surprisingly strong absorption near 3 \mum \citep{Baines2011epsc} that
was not seen anywhere else on Saturn (Fig. 2). Ammonia ice (\nhtx) and
ammonium hydrosulfide (\nhfshx), as well as water ice, are possible
contributors because they are all strong absorbers in this region of
the spectrum.
While all of these molecules are also expected to produce clouds on
Saturn, a thick overlying haze has largely shielded them from view.
Until recently, there has been little spectroscopic evidence of any of
them except in association with small storm clouds in the southern
hemisphere, where excess absorption near 2.7 \mum suggested the
presence of freshly condensed \nht \citep{Baines2009stormclouds}.  The
Great Storm of 2010-2011 is by far the largest 3-\mum absorbing
feature ever seen on Saturn and provides the best opportunity for
detailed spectroscopic analysis.

Our objective here is to model the composition and vertical structure
of the storm feature, using detailed spectral characteristics to
determine the dominant components of the storm cloud particles, as
well as to estimate the size of the particles and whether they are
likely to be composite or pure.  Of all the pure substances we tried,
particles of \nht provided the best fit to the absorption feature.
However, pure \nht presents problems because its narrow strong
absorption feature at 2.96 \mumx, which should be resolved by VIMS, is
seen only weakly if at all.  Other characteristics of \nht do appear
to fit well, but some mechanism that suppresses the 2.96-\mum feature
seems to be required.  Contributions of other absorbers such as \nhfsh
and water ice can provide improved fits, suggesting that all three
substances might be present in the cloud particles, with \nht being
the most prominent contributor and water the most probable secondary
component.  In the following we first present the VIMS observations,
describe our spectral analysis techniques, evaluate spectral fits for
pure substances, then consider alternatives involving combinations of
\nht with other substances, both as co-mingled pure particles and as
composite particles, and also as layers of different compositions.

\section{Observations.}

\begin{figure*}[!htb]\centering
\hspace{-0.15in}\includegraphics[width=6in]{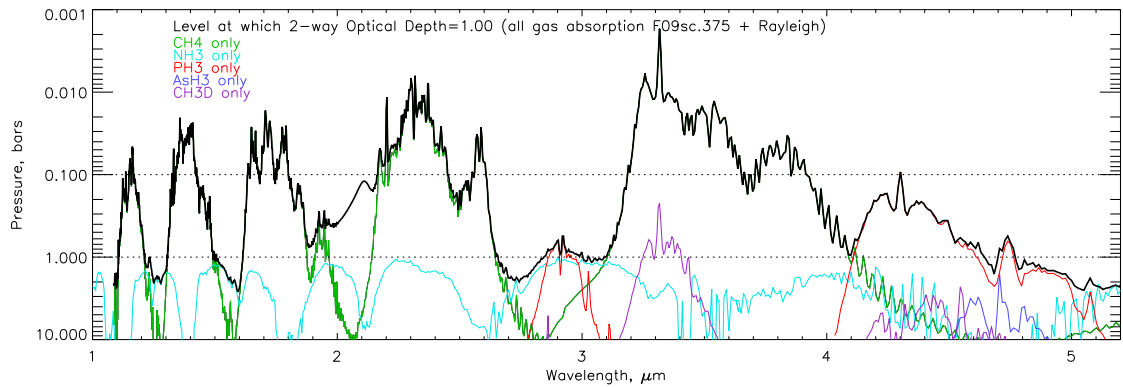}
\caption{Penetration depth of near-IR photons indicated by pressures
  at which a unit albedo reflecting layer produces an external I/F of
  1/e at normal incidence and viewing. Left: the CCD spectral region.
  Curves are shown for methane only (green), ammonia only (cyan),
  phosphine only (red), arsine only (blue), CH$_3$D only (purple), and
  all gases combined (black), assuming the \cite{Lindal1985}
  temperature profile, the He/H$_2$ ratio of \cite{Conrath2000}, the
  ammonia profile of \cite{Prinn1984}, and the phosphine profile of
  \cite{Fletcher2009ph3} with a scale height of 0.375 times the gas
  scale height. Included, but not shown separately, is H$_2$ CIA,
  which peaks near 2.1 \mumx.}
\label{Fig:pendepth}
\end{figure*}

We used publicly
available spectra acquired by VIMS during 24 February 2011 at 00:36:36
UT, data cube number 1677201862\_3, acquired under observation name
VIMS\_145SA\_WIND5HR001, at a phase angle of 52\deg and a range of
1.84$\times 10^6$ km.  The VIMS instrument and investigation are
described by \cite{Miller1996SPIE} and \cite{Brown2004SSR}. The
instrument's spectral range is 0.35-5.1 $\mu$m, with an effective
pixel size of 0.5 milliradian on a side and a near-IR spectral
resolution of approximately 15 nm (sampled at intervals of
approximately 16 nm).  The atmospheric pressures that can be sampled
by this wide spectral range are indicated in Fig.\ \ref{Fig:pendepth},
which was computed for a clear atmosphere and thus over-estimates the
penetration depths below the 300 mb level, where significant aerosol
opacity is a limiting factor.  Methane is the most prominent absorbing
gas at most wavelengths.  Hydrogen collision-induced absorption is
dominant in the 1.95-2.15 \mum range.  Phosphine is prominent in the
2.8-3 \mum and  4.2-5 \mum regions.  Arsine and \chtd have
much smaller effects, and germane has too small an effect to be
noticed in our investigation.  

\begin{figure}[!htb]\centering
\hspace{-0.0in}\includegraphics[width=3in]{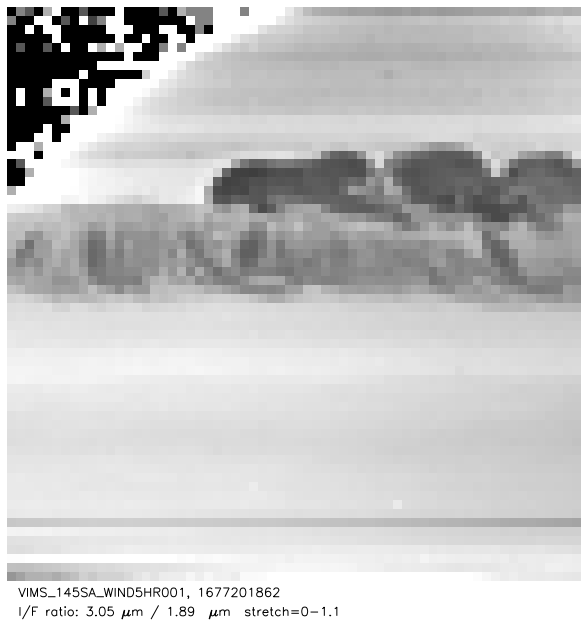}
\caption{VIMS images of I/F ratio between 3.05 \mum (numerator) and
  1.89 \mum (denominator).  Absorption by NH$_3$ is greatest where the
  ratio is smallest.}
\label{Fig:ratios}
\end{figure}

A selection of these earliest VIMS observations of the Great Storm are
shown in Fig. \ref{Fig:VIMSimages}.  Remapped images are shown at
three key wavelengths and spectra from seven representative locations
are displayed, including means of two just upstream of the storm head
and two within the storm head itself.  Compared to spectra at other
locations, the cloud reflectivities in the storm head are the highest at
4.081 \mum and the lowest at 3.048 \mumx.  As shown in
Fig.\ \ref{Fig:VIMSimages}E, spectra obtained upstream (locations 3 \&
4) display absorption between 2.8 and 3 \mum that is due to phosphine,
but show little absorption at 3.05 \mumx.
The upstream clouds are also more transparent at thermal wavelengths,
evident from their large apparent I/F values near 5 \mumx. The storm
particles are brighter than the upstream cloud particles in most
window regions, somewhat more absorbing at 2.7 \mum, but dramatically
more absorbing at 3.048 \mumx.  The image at 1.887 \mum
(Fig.\ \ref{Fig:VIMSimages}A) is used for the green component of the
composite image because it provides the closest match to the I/F at
3.05 \mum for those regions that do not contain a 3-\mum particle
absorption, both of which, along with 4.081 \mumx, are pseudo
continuum wavelengths that are relatively unaffected by gas
absorption.  This means that the ratio of I/F at 3.048 \mum to that at
1.887 \mum should be a good measure of the degree of 3-\mum
absorption, as illustrated in Fig.\ \ref{Fig:ratios}.  The regions of
high reflectivity at 4.081 \mum are places where the cloud particles
are relatively large (probably 1 \mum or more), and strong absorption
at 3.048 \mum (Fig.\ \ref{Fig:VIMSimages}B) appears as bright orange
in the color composite (Fig.\ \ref{Fig:VIMSimages}D).

\section{VIMS instrument characteristics and corrections}

Before proceeding with the quantitative analysis of the VIMS spectra, we
first describe special characteristics of the VIMS instrument and summarize how
we handle data processing, photometry, and navigation.

\subsection{VIMS responsivity errors and corrections}\label{Sec:artifact}
The responsivity of the VIMS detector array is significantly reduced
at joints between order sorting filters, which occur at wavelengths of
1.64 \mumx, 2.98 \mumx, and 3.85 \mum
\citep{Miller1996SPIE,Brown2004SSR}.  The decline is most severe at
1.64 \mum and 3.85 \mumx, and VIMS spectra in these regions cannot be
trusted, as shown by \cite{Sro2010vims} for the 1.64 \mum joint, and
by our own radiation transfer modeling for the 3.85 \mum joint.  The
spectra near the 2.98-\mum joint are reasonably well corrected in
spectra obtained during the Jupiter encounter, as shown by
\cite{Sro2010vims}, but minor problems are present in the VIMS spectra
of Saturn. The 2.98-\mum joint is unfortunately located so close to
the prominent \nht ice feature near 2.96-\mum that its effects can be
misinterpreted as being due to the presence of \nht ice.  In fact, all
the raw VIMS spectra of Saturn do contain a dip very similar to what
might be produced by a small component of \nht ice in the main
tropospheric haze layer. However, the spatial variation of the feature
is much more easily understood as an instrumental artifact. The
feature appears to be a fractional effect with no dependence on
latitude or view angle, but only on the x-coordinate of the VIMS
spectral data cube. It is not plausible that an atmospheric effect
could produce this characteristic. Our interpretation is confirmed by
ISO measurements of Saturn's 3-\mum spectrum by \cite{Encrenaz1999},
which provide no evidence of such a feature.

To correct for the responsivity error caused by the 2.98-\mum filter joint 
we collected spectra covering regions without
strong convective features, then interpolated across the artifact for
each spectrum and used the fractional depression below the interpolated line as a measure of the
responsivity error that needed to be corrected.  We found essentially
no variation along image columns (y-direction), and thus used an average fractional depression
along each column to define the correction for any position along a
column.  We did find a significant variation perpendicular to the
column direction, however.  The fractional error for band 127 varied
smoothly from 0.75 at column zero to 0.52 at column 63, a change that
is about 30 times the standard deviation along that column.  A physical
explanation for this behavior in terms of instrument design and
operation remains to be determined.  With this correction we are able
to separate the effects of the filter joint from those of true
particulate absorption in the same region of the spectrum.

\subsection{VIMS errors at low signal level}\label{Sec:lowsig}

Although a specific cause has not been identified, \cite{Sro2010vims}
showed that very dark regions of VIMS spectra tend to be significantly
brighter than independent observations from the ground
and from HST using NICMOS.  This occurs at very low signal levels,
for which the instrument is responding with 1 Digital Number or less.
What appears to be a small offset, just a fraction of a DN, if
not corrected for, leads to excessive stratospheric
haze opacity in models designed to match these spectra.  As we have
no well-defined correction for this effect, it limits our ability
to constrain the stratospheric haze, which is only noticed at wavelengths
for which the atmosphere is extremely dark because of strong gas absorption.

\subsection{VIMS wavelength corrections}

\begin{figure}[!hbt]\centering
\includegraphics[width=3.25in]{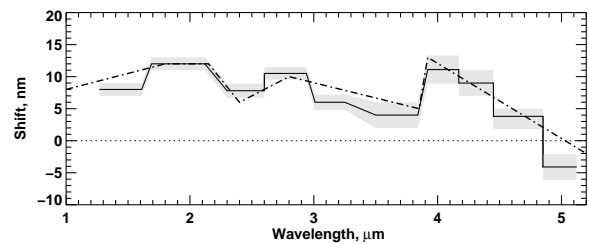}
\caption{VIMS wavelength shifts obtained from matching individual
  spectral bands within 10 segments (solid stair-step with gray error
  bands), and adopted ramp shifts (dot-dashed lines) that provide
  improved matches to complete spectra.}
\label{Fig:wavecal}
\end{figure}

The wavelength calibration that worked well for modeling VIMS Jupiter
spectra \citep{Sro2010vims} came from PDS example calibrated data
cubes, which used a set of wavelengths that was different from those
contained in the Jupiter flyby data cube headers. According to
\cite{McCord2004}, the VIMS wavelength calibration changed since
launch by about 12 to 22 nm. Using comparisons with NIMS, they revised
the VIMS wavelength calibration to what we believe is in the PDS
example calibrated data cubes. These wavelengths are the same as
appear in current PDS data cube headers for Saturn (except for the
last wavelength).  They also match what \cite{Cruikshank2010} refer to
as the RC15 wavelength calibration.  But, while these wavelengths
worked well for the December 2000 Jupiter flyby data,
hen used with the 2010-2011 Saturn data cubes, they produce conflicts
between measurements and model calculations regarding the positions of
gaseous absorption bands. To align VIMS spectra with these absorption
bands, which are accurately known from line-by-line calculations,
requires shifts of the VIMS wavelengths that vary with wavelength.  In
Fig.\ \ref{Fig:wavecal}, we show the results of matching 10
wavelength segments containing different absorption bands.  The
required shifts are a significant fraction of the 15-nm VIMS
resolution, ranging from 12 nm near 2 \mum to -4 nm near 5 \mumx.  We
found that ramped wavelength shifts lead to better spectral fits than
constant segments.  The ramp functions we adopted, shown as
dot-dashed lines, were adjusted to roughly follow the segment shifts, and
to cross zero near 5 \mumx, where the \cite{Fletcher2011vims} analysis
of night-side Saturn spectra from 2006 found good agreement between
VIMS and model wavelengths.
As shown in Section \ref{Sec:samplefit}, fit quality is much worse if
unshifted wavelengths are used.  
This implies that the VIMS instrument changed behavior between the
early part of the 2000 Jupiter encounter and operations at Saturn in
2010-2011, by almost as much as the change from pre-launch to the time
of the Jupiter approach.  
Whether the change in
instrument temperatures between these two time periods could have
been a factor remains to be determined.

\subsection{Image processing and navigation}

VIMS image cubes were processed using the pipeline processing code
downloaded from the Planetary Data System, described by
\cite{McCord2004}.  
To navigate the images, we extracted spacecraft and target body states, and
instrument pointing information from the index.tab files contained in
the PDS archive volumes.  The sun to target distance was retrieved
from Cassini NAIF SP kernel files, bypassing the VIMS pipeline code, which used a
fixed distance (9.033 AU). These data allow accurate transformation between
image coordinates and planet coordinates using SPICE
toolkit software \citep{Acton1996}. We processed the VIMS data cube up
through the ``specific energy'' product stage using pipeline software,
and then computed the I/F using the correct sun-target distance and
the solar spectrum provided in the PDS volumes. The VIMS calibration
uncertainty 
is at least as large as the near-IR solar spectral irradiance
uncertainty of $\sim$3\%-5\% \citep{Colina1996}.  It appears to be
less than 10\% uncertain, based on window-region comparisons made by
\cite{Sro2010vims} with independent spectra of Jupiter.

\subsection{VIMS noise characteristics.}\label{Sec:vimsnoise}

For most applications at moderate signal levels, the random noise
level of VIMS is very small, nominally less than 1 DN.
\cite{Sro2010vims} tried to better characterize the
noise by comparing spectra in cubes that imaged the same region of
Jupiter with as small a time difference as possible.  
They found measurements in the same channel to be within about 1.2 DN
RMS for a single measurement, even for signal levels up to 2000 DN or
more.  However, at signals 100 times smaller, as found near 2.3 \mumx,
\cite{Sro2010vims} found a much larger fractional noise, which can be
characterized as an I/F offset noise, which they crudely estimated as
$\sim$5$\times 10^{-4}$ from the on-disk spectra.
Other possible sources of uncertainty, though not strictly random
noise, are scattered light inside the spectrometer, wavelength errors,
line-spread uncertainties, and wavelength-dependent absolute calibration errors. 

\section{Radiation transfer calculations}

\subsection{Atmospheric structure and composition}
\begin{figure}[!hbt]\centering
\includegraphics[width=3.25in]{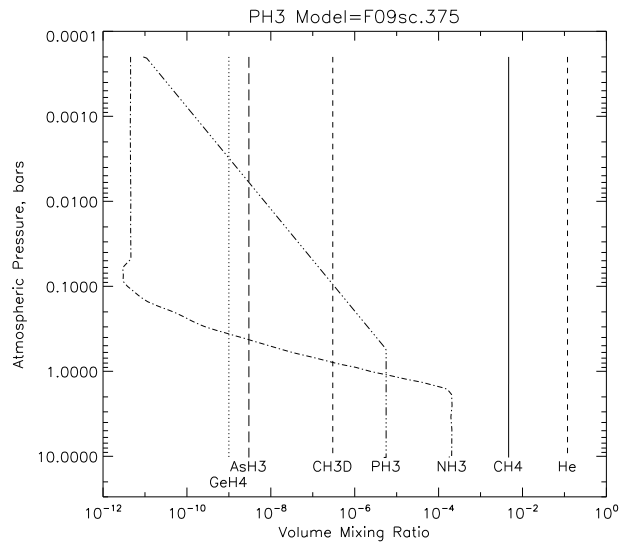}
\caption{Assumed volume mixing ratios of spectroscopically important gases
in the atmosphere of Saturn. See text for references.}
\label{Fig:gasmix}
\end{figure}

\vspace{-0.05in} The assumed composition of the atmosphere as a
function of pressure is displayed in Fig.\ \ref{Fig:gasmix}. We used
the \cite{Lindal1985} temperature structure between 0.2 mb and 1.3
bars and dry adiabatic extrapolation to approximate the
structure at deeper levels.  \cite{Lindal1985} used a He/H$_2$ ratio
of 0.06/0.94 in deriving their profile, and noted that the profile
could be scaled to account for a different He/H$_2$ ratio.  We used
the revised ratio of 0.135$\pm$0.025 due to \cite{Conrath2000}, but
did not scale the Lindal et al. thermal profile because that would
result in disagreement with the thermal emission spectrum. The
\cite{Conrath2000} revision was motivated by a disagreement between
the pre-Galileo He/H$_2$ ratio for Jupiter and the accurate in situ
measurements by the Galileo Probe. The pre-Galileo result was based on
a combined analysis of IRIS spectra and the Voyager-measured radio
occultation refractivity profile.  In an attempt to avoid whatever
error had occurred in that Jupiter analysis, Conrath and Gautier based
their Saturn revision entirely on Voyager IRIS observations.  A lower
value of 0.08 was given by \cite{Fouchet2009} based on Cassini CIRS
spectra and Cassini radio occultations analyzed by
\cite{Gautier2006cosp}. Fortunately, these uncertainties in the
He/H$_2$ ratio have little effect on our analysis.  As long as the
derived thermal profiles remain in agreement with the thermal emission
spectra, which is true for all of these results, we found that our
derived cloud model parameters are not very sensitive to the He/H$_2$
ratio. Test cases for ratios from 0.06 to 0.135 show that the derived
cloud parameters don't change more than their uncertainty limits.
After trying a variety of values for the \chf volume mixing ratio
(VMR), we found the best fits were generally obtained with the
\cite{Fletcher2009ch4saturn} value (4.7$\pm$0.2)$\times 10^{-3}$,
which corresponds to a \chfx/H$_2$ ratio of 5.3$\times 10^{-3}$.  For
\chtd we also used the \cite{Fletcher2009ch4saturn} VMR value of
3$\times 10^{-7}$.
The most important variable gas is PH$_3$ and its vertical profile
needs to be adjusted to fit VIMS spectra. \cite{Fletcher2009ph3} has
shown that PH$_3$ is enhanced over the equator and depleted over belts
and within hot cyclonic polar vortices. We found that a reasonably
effective parameterization was a fixed deep mixing ratio and an
adjustable scale height defining its rate of decrease for pressures
less than about 600 mb. Our best fits outside the storm region were
obtained with a deep mixing ratio of 6.4$\times 10^{-6}$, which is the
global mean value of \cite{Fletcher2009ph3}, and a scale height of
37.5\% of the pressure scale height.  Inside the storm region better
fits were obtained with a more extended \pht profile using the same
deep mixing ratio and 50\% of the pressure scale height. The greater scale
height within the storm is consistent with increased upward transport
of \phtx, as expected in regions of strong convection.  We used the
\nht profile given by \cite{Prinn1984}, which has a constant deep
mixing ratio of 2.05$\times 10^{-4}$, based on
\cite{Courtin1984}. More recent VLA results include a value of (4.8
$\pm$1)$\times 10^{-4}$ by \cite{Briggs1989} and some evidence for
depletion of \nht in the 2-4 bar region, which they interpret as
evidence for an \nhfsh cloud and H$_2$S vapor. Arsine has a noticeable
affect on the VIMS spectra near 5 \mumx, which is where ammonia gas
also plays a relatively minor role. Although \nht absorbs over a wide
wavelength range, its effects are largely shielded by overlying cloud
opacity.  It does however have a minor influence in blocking some of
the thermal radiation at wavelengths beyond 4.6 \mumx.

\subsection{Gas absorption models}
\vspace{-0.05in} Because we wanted our correlated-k methane absorption
models to be based on line-by-line calculations, we limited our
spectral analysis to wavelengths greater than 1.268 \mumx. Recent
advances
in measuring line spectra \citep{Campargue2012refine} have improved
line-by-line calculations so much that for $\lambda >$ 1.268 \mum they
are now likely more accurate than band models (see \cite{Sro2012LBL}
for comparisons and correlated-k models).  We used the same line data
for computing \chtd absorption models.  For \nht we used the combined
correlated-k absorption model described by \cite{Sro2010iso}, which is
based primarily on the Goody-Lorentz band model of \cite{Bowles2008}.
To model phosphine (\phtx) absorption we used our own exponential sum
approximations based on the line data of \cite{Butler2006} in the
2.8-3.1 \mum region and on the \cite{Rothman2009} HITRAN 2008 line
data in the 4.1-5.1 \mum region. Our model of AsH$_3$ absorption is
based on line data from \cite{Tarrago1996} (via G. Bjoraker, via
B. B\'ezard, personal communication).  Where \chf and \pht gas
absorptions overlap we compute opacities for 100 combinations of 10
\chf terms by 10 \pht terms and then sort and refit to a 10-term
weighted sum, following the suggestion of \cite{Lacis1991}. We do the
same with all other gas overlaps in succession, which reduces all gas
absorptions to ten-term exponential sum models at all wavelengths,
even in regions with overlap of several gases.

Collision-induced absorption (CIA) for H$_2$ and H$_2$-He was
calculated using programs downloaded from the Atmospheres Node of the
Planetary Data System, which are documented by \cite{Borysow1991h2h2f,
  Borysow1993errat} for the H$_2$-H$_2$ fundamental band,
\cite{Zheng1995h2h2o1} for the first H$_2$-H$_2$ overtone band, and by
\cite{Borysow1992h2he} for H$_2$-He bands. Although these programs
and/or tables generally provide coefficients only for normal and
equilibrium hydrogen, absorption for any para fraction can be derived
by appropriate linear combinations of these values
\citep{Birnbaum1996}.  We used equilibrium hydrogen generally and made test
calculations with normal hydrogen to assure ourselves that our conclusions
were not sensitive to the ortho/para ratio.

\subsection{Multiple scattering methods}

For multiple scattering calculations we make use of a modification of
the doubling and adding code described by \cite{ Sro2005raman,
  Sro2005pol}.  The modification is to remove the Raman source term,
which is not needed at near-IR wavelengths, and restore the original
blackbody emission source term \citep{Evans1991} to handle the 5-\mum
region of the spectrum.  We used a grid of 44 pressure levels from 0.5
mb to 10 bars, distributed roughly in equal log increments, except
that additional layers are introduced where cloud layers are
inserted. As a function of wavelength we stop adding layers after the
one-way vertical gas optical depth exceeds six.
To model the medium phase angle VIMS observations
we used 16 quadrature points in both zenith angle and azimuth.
We followed \cite{Sro2010iso, Sro2010vims} in modeling
coated-sphere particle scattering using code originated by
\cite{Toon1981}. 
We approximated the line-spread function of the VIMS instrument as a
Gaussian of FWHM = 0.015 $\mu$m, then collected all the opacity values
within $\pm$FWHM of the sample wavelength, weighted those according to
the relative amplitude of the line-spread function, then sorted and
refit to ten terms again.  
A special treatment is required in the 2.9-3.0 \mum
region where \nht cloud particles have very sharp absorption features.
In this region model calculations were made at 5-\icm intervals (the
maximum sampling frequency of our correlated-k models) and
subsequently smoothed to VIMS resolution.

\subsection{Fitting cloud models to observations}

Our cloud models usually included both vertically extended layers,
where layer boundaries could be constrained by the observations, and
compact layers, where the vertical extent of cloud layers could not be
constrained.  In some cases we evaluate both compact and vertically extended
models, often finding that both can work for layers that are weakly
constrained.  For models where the number of adjustable parameters is
relatively small, parameters were adjusted to minimize $\chi^2$ using
a form of the Levenberg-Marquardt algorithm, as described by
\cite{Press1992}.
In most cases, model complexity is too large to reach automated
convergence of all parameters in a practical time period (or ever).
To reduce the scale of the fitting problem and increase algorithm
stability, we use constraining assumptions for a subset of the
parameters, which usually cannot be assigned reliable uncertainty
estimates.  It is important to note that formal uncertainties in the
remaining fitted parameters are only valid within the context of the
constraining assumptions and cannot be treated as absolute
uncertainties.

To speed the fitting process we reduce the total number of spectral
comparisons to a set of 48 wavelengths that sample key spectral
features and a wide range of gaseous absorption strengths.  After
fitting adjustable parameters to minimize \chisq for the sampled
points, we then compute a complete spectrum for more detailed
comparisons. To test for the presence of bias due to the specific
wavelengths sampled in constraining our fits, we did a fit to the
Location 4 spectrum with the complete set of 225 uniformly spaced
wavelengths as constraints.  The two resulting sets of fit parameters
agreed within their uncertainties, indicating no significant bias.

\subsection{Estimating uncertainties in derived model parameters}

The constraining properties of the observations depend on the
nature of measurement errors as well as on the errors in the
modeling of atmospheric opacity.  
As previously noted, the VIMS random I/F measurement errors in a given
wavelength channel are negligible except at very low signal
levels. Systematic errors are more of a concern, both in absolute
calibration and in the wavelength scale, which can itself result in
I/F errors when comparing to models at slightly incorrect wavelengths.
A more important error source, which is not strictly random, but can
vary from wavelength to wavelength, is the uncertainty in radiation
transfer modeling due to the uncertainty in absorption coefficient
models. The
transmission curves on which previous methane opacity models were
based had an assumed error of 0.05 \citep{Irwin2006ch42e}.  Although
not well characterized, the correlated-k models derived from
line-by-line calculations also produce uncertainties.

 To try to account for these effects we used the following procedure.
 Initial model fits are used to establish a rough characterization of
 the vertical opacity structure of the atmosphere.  We then calculate
 model spectra for the rough model, and for two perturbations: (1)
 optical depths offset by 0.01 and (2) optical depths increased by
 10\%.  The resulting I/F differences are root sum squared with I/F
 errors due to wavelength uncertainties, and with an I/F offset
 uncertainty of 5$\times 10^{-4}$ 
and a measurement/relative calibration uncertainty assumed to be 1\%.
The I/F error associated with wavelength uncertainty is calculated
from the derivative of the I/F spectrum with respect to wavelength
times an estimated wavelength uncertainty of 0.002 \mumx, which is 1/8
of a VIMS line width.  In some cases we assume a simpler uncertainty
model, which is not cloud model dependent, but is characterized by an
I/F offset uncertainty of 0.002 and a fractional I/F uncertainty of
10\%. The expected $\chi^2$ value for the best fit is equal to the
number of degrees of freedom $N_F$ (number of points of comparison
minus the number of adjusted parameters) and the expected uncertainty
in the value is equal to $\sqrt{2N_F}$ \citep{Rice1995}.  However,
both error models result in minimum $\chi^2$ values several times
$N_F$. This means that either our error estimates are too low, or our
physical models of the clouds are incomplete.  There are many reasons
for the latter to be true, including not-accounted-for effects of
non-spherical particles, multi-coated particles (more than two
compositional layers), and heterogeneous structures, as well as more
complex vertical distributions. To approximately correct for bad error
estimates and incomplete physics, we rescaled our \chisq values by the
factor $N_F/\chi^2_{MIN}$ before finding confidence limits. Because
higher than expected \chisq values are probably not simply due to bad
error estimates, this procedure is not entirely valid, but is clearly
better than making no adjustment.

\section{Modeling cloud structure outside the storm regions.}

\subsection{Sample fits using a simplified cloud structure}\label{Sec:samplefit}

Near IR spectra outside the storm regions can be very well fit by
relatively simple cloud structures, which are similar to those used
successfully to fit spectra in the CCD range from 0.3 to 1 \mum
\citep{Munoz2004,Kark2005Icar,Perez-Hoyos2005}.  These structures
contain a vertically diffuse stratospheric haze, an optically thick
and physically thick main cloud layer, which requires visible and UV
absorption, but can fit near IR spectra with conservative cloud
particles, and a deeper layer, to which most wavelengths have little
sensitivity, but which can be constrained by the thermal emission that
can be sensed in the 5-\mum region.  Fits of such a model structure to
the VIMS spectra from all non-storm locations identified in
Figs. \ref{Fig:VIMSimages} are summarized in Table \ref{Tbl:parvals},
where parameter values, uncertainties, and \chisq values are all
derived from the subset of 48 spectral samples discussed
previously. In all these fits we characterized the deep cloud as
physically compact, but optically thick (arbitrarily set to $\tau$ =
30) and absorbing (set to $\varpi$=0.9 to provide adequate blocking of
thermal emission).  The only adjustable parameter of this layer was
its effective pressure. As will be shown in the following, other deep
cloud options could also be made to work.  Cloud particles in the main
haze layer (generally between 120 and 700 mb) were treated as Mie
particles with refractive index n=1.4+0i. We fit the upper and lower pressures of this
layer and its optical depth at 2 \mumx, and within the boundaries of
this layer we assumed a vertical scale height equal to the gas scale
height. Because our \chisq values are several times the expected value
(the number of degrees of freedom), it is clear that either our model
lacks needed physics or our measurement error estimate is too small, 
by a factor of $\sqrt{\chi^2_{MIN}/N_{F}}$, or some of both.  If we
assume that our error estimate is the problem, then we need to
multiply our standard parameter uncertainty estimate (the change
needed to produce $\Delta \chi^2 =1$) by the same factor of
$\sqrt{\chi^2_{MIN}/N_{F}}$.  Error estimates in Table
\ref{Tbl:parvals} have this correction applied.

The complete best-fit model spectrum for location 4 is displayed in Fig.\ \ref{Fig:fitsample}.
We also evaluated the fit quality of this spectrum using
$\chi^2$ computed for all measured points, assuming a simplistic random
error model with a reflectivity (I/F) fractional uncertainty
of 0.1 and an offset uncertainty of 0.002, which provides more of a
relative than absolute judgment of fit quality.  This provides a convenient
way to compare full resolution versions of the best-fit model spectra.  Two \chisq values are
shown in the legend of Fig.\ \ref{Fig:fitsample}: the one labeled $\chi^2/N$(TOT) is
computed for 1.268-5.15 \mum wavelength range, and the one labeled $\chi^2/N$(ABS) is
computed for the 2.65-3.2 \mum range where particulate absorption
features can be prominent (in the storm region).  All are computed with respect to the
wavelength-shifted spectrum.  If computed relative to an unshifted
spectrum the \chisq/N (ABS) increases from 0.26 to 2.3 and \chisq/N
(TOT) increases from 0.98 to 3.8.  This is a dramatic change that
clearly confirms the need for wavelength corrections.  Although the
fit quality for this model is very high, the model is not unique.  A
deep cloud that is vertically extended, with a constant optical depth
per unit pressure, can fit just as well, provided it contains
some absorption that increases with wavelength.

\subsection{Sensitivity of model spectrum to model parameters}

\begin{figure*}[!hbt]\centering
\includegraphics[width=6.in]{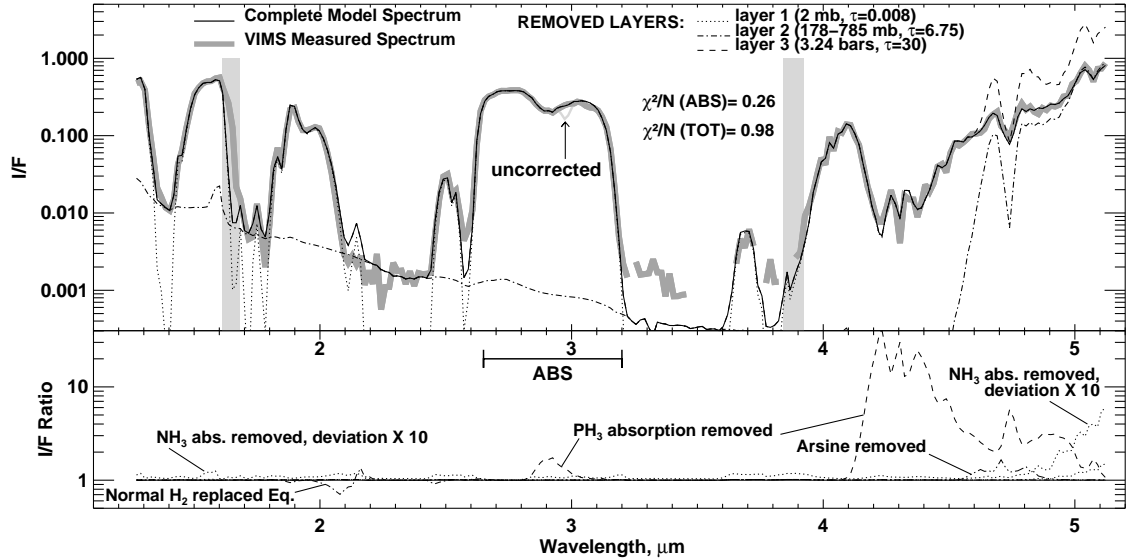}
\caption{Example model spectrum (thin solid curve) fit to the VIMS
  mid-latitude spectrum (thick gray curve) from location 4 in
  Fig.\ \ref{Fig:VIMSimages}.  The model parameters are given in
  Table\ \ref{Tbl:parvals}. Contributions from each layer can be
  inferred from the spectral changes produced by removing it.
  Relative fit quality for the entire spectrum (excluding the two
  regions indicated by vertical gray bars) is indicated by the
  $\chi^2$/N (TOT) values, and for the 2.65 \mum - 3.2 \mum region of
  particulate absorption by $\chi^2$/N (ABS) values, where there are
  33 points in the ABS region and 223 in total. Sensitivities to gas
  absorptions by \nht (dotted), \pht (solid), AsH$_3$ (double
  dot-dash) and normal vs. equilibrium H$_2$ (dot-dash) are indicated
  by ratio spectra in the bottom panel (modified model/best-fit
  model). Greater sensitivity to these gases would be obtained in
  regions of lower cloud opacity.  The importance of thermal emission
  in constraining deep cloud opacity can be seen from the large I/F
  increase caused by removal of layer 3. Vertical gray bars indicate
  where order sorting filter joints cause uncorrected VIMS I/F
  uncertainties. At the joint near 2.96 \mumx, the measured spectrum
  has been corrected for responsivity errors following a procedure
  described in the text. The thin lighter gray curve shows the
  original uncorrected spectrum.  The good agreement of the corrected
  spectrum with model calculations in this region attest to the
  efficacy of the correction.}
\label{Fig:fitsample}
\end{figure*}

Figure\ \ref{Fig:fitsample} also displays the effect of removing each
model layer one at a time, and also the effect of removing various
gases from the atmospheric absorption model. Next to methane,
phosphine has the most significant effect on the spectrum. Hydrogen
CIA is next most significant, but limited to wavelengths from 2 to 2.2
\mumx.  The stratospheric haze provides the baseline I/F in the
darkest parts of the spectrum.  However, because of the anomalously
large I/F values measured by VIMS at very small I/F levels
\citep{Sro2010vims}, and the small reflectivity of the small
stratospheric haze particles at near IR wavelengths, what can be
learned about that haze from the VIMS observations is rather limited.
We found that a Mie particle radius of 0.1 \mum or a Henyey-Greenstein
scatterer with wavelength dependence of $\lambda^{-3.5}$, a pressure
level of 2 mb, and an adjustable optical depth provides a suitable
model. But this is not a very realistic physical model of the
stratospheric haze. Extending the haze downward as prior CCD-based
models have done, does not improve the accuracy of fits here, probably
due to the offset errors of the VIMS instrument.

\subsection{Alternate vertical distributions}

\begin{figure*}[!htb]\centering
\includegraphics[width=5.25in]{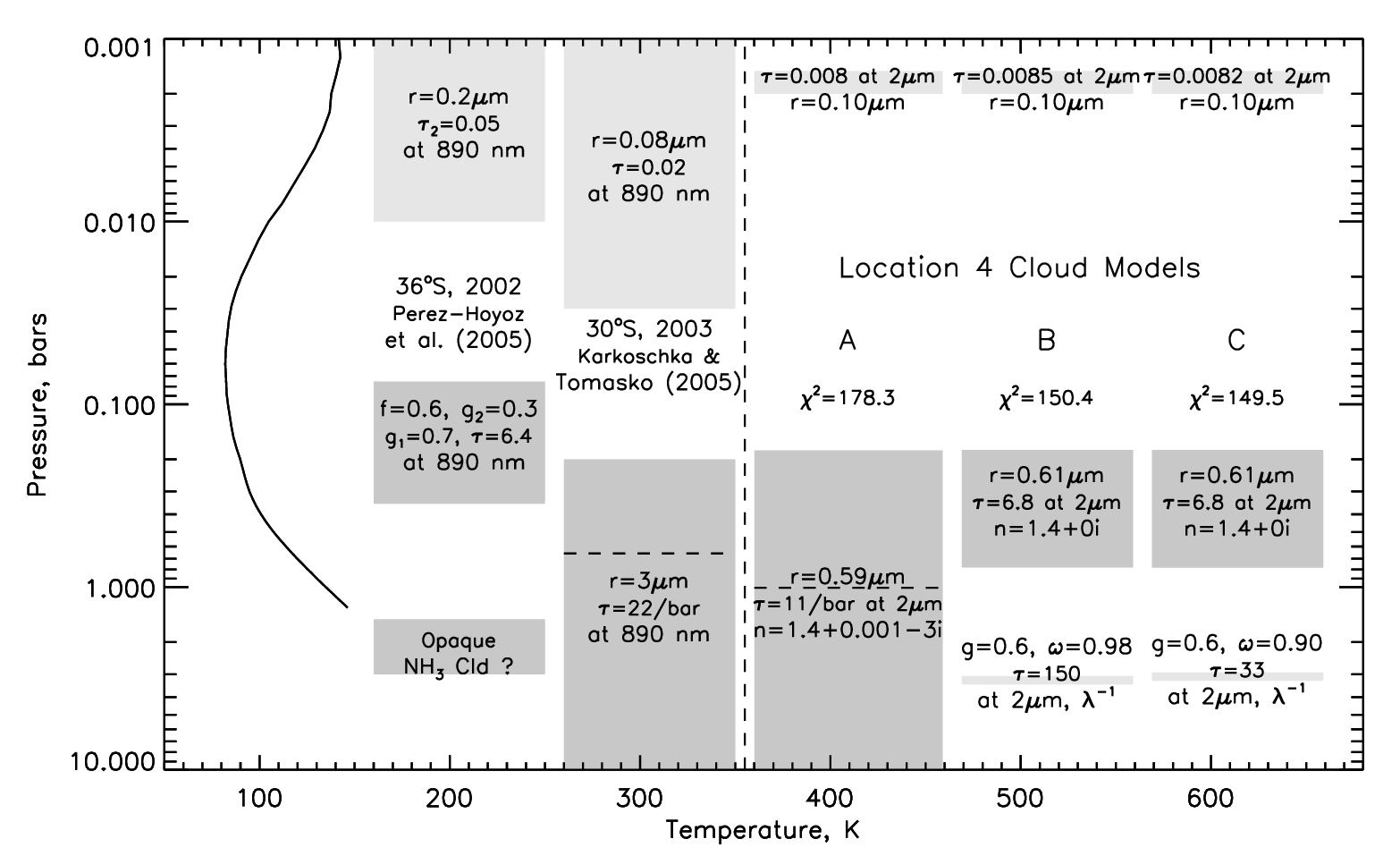}\\
\caption{Sample cloud structure models for Saturn according to
  \cite{Perez-Hoyos2005} and \cite{Kark2005Icar} and, to the right of
  the vertical dashed line, our model fits to the spectrum from location
4 in Fig.\ \ref{Fig:VIMSimages}. Structure A uses a uses a uniformly mixed
single particle type from cloud top to the bottom of the atmosphere,
as used by \cite{Kark2005Icar}. To provide sufficient absorption of 
long wavelengths this requires an imaginary index that increases 
with wavelength.  Structures B and C, which are similar to that
used by \cite{Perez-Hoyos2005},  differ in the single-scattering
albedo assumed for the lower cloud layer. Higher albedo requires
larger optical depth to block thermal emissions. \chisq values shown
are for fits to the 48-wavelength sampled spectra.}
\label{Fig:cartoon1}
\end{figure*}

The variety of vertical distributions that can fit the near IR spectra
is similar to the variety that have been used to fit the visible
spectra, as illustrated in Fig.\ \ref{Fig:cartoon1}, where we compare
our fits to the non-storm spectrum from location 4 in
Fig.\ \ref{Fig:VIMSimages} to models of \cite{Perez-Hoyos2005} and
\cite{Kark2005Icar} applicable to similar latitudes, but different
time periods.  The disagreements are not surprising, given the secular
variation that might have occurred between the relevant observing
times.  We tried both broadly diffuse vertical distributions suggested
by \cite{Kark2005Icar} and the more compact models used by
\cite{Perez-Hoyos2005}.  We found reasonably good fits were possible
with both types of models, even to the 5-\mum region.  By making use
of thermal wavelengths ($\lambda >$ 4.8 \mumx), we were able to place
better constraints on the deep cloud structure than were the
previously cited authors, though here again there is not a clear
distinction that can be made between compact and diffuse models.  If
we assume that the upper cloud has a real index of 1.4, and that the
deep cloud is nearly conservative ($\varpi$=0.98) and compact (Model B
in Fig.\ \ref{Fig:cartoon1}), then we find that the cloud needs to
reside near 3.3 bars and have a high optical depth ($\tau \sim
150$). But if the cloud is more strongly absorbing ($\varpi$=0.9 as in
Model C in Fig.\ \ref{Fig:cartoon1}), then the optical depth
requirement is greatly reduced ($\tau \sim 30$) while the pressure
remains essentially unchanged.  In all these cases the main haze layer
visible from above was assumed to have a refractive real index of 1.4,
to be uniformly mixed with gas, and to have adjustable top and bottom
pressures, which reached typical values of 178 mb and 780 mb
respectively, while the optical depth did not deviate much from 6.8 at
2 \mumx.  Although the compact lower cloud models with a separated
upper tropospheric layer (extending from 178 mb to 780 mb) provided
the best fits (B and C in Fig.\ \ref{Fig:cartoon1}), we were also able
to get reasonable fits with a semi-infinite diffuse cloud extending
downward from 178 mb, using an opacity of 11 optical depths/bar (A in
Fig.\ \ref{Fig:cartoon1}), with a real index of 1.4 and an imaginary
index linearly increasing from 0.001 at 1 \mum to 0.003 at 5.5
\mumx. This latter model is similar to that used by
\cite{Kark2005Icar}. Perhaps the most significant deviation is that
our VIMS-based fits for the stratospheric haze layer are not
consistent with that haze extending down to 100-300 mb, as found by
prior fits.  We believe that this disagreement arises from an artifact
of the VIMS instrument in which very low I/F values are too high, as
explained in Section \ref{Sec:lowsig}.

\subsection{Latitudinal variations in non-storm regions}

\begin{figure}[!htb]\centering
\hspace{-0.05in}\includegraphics[width=3.5in]{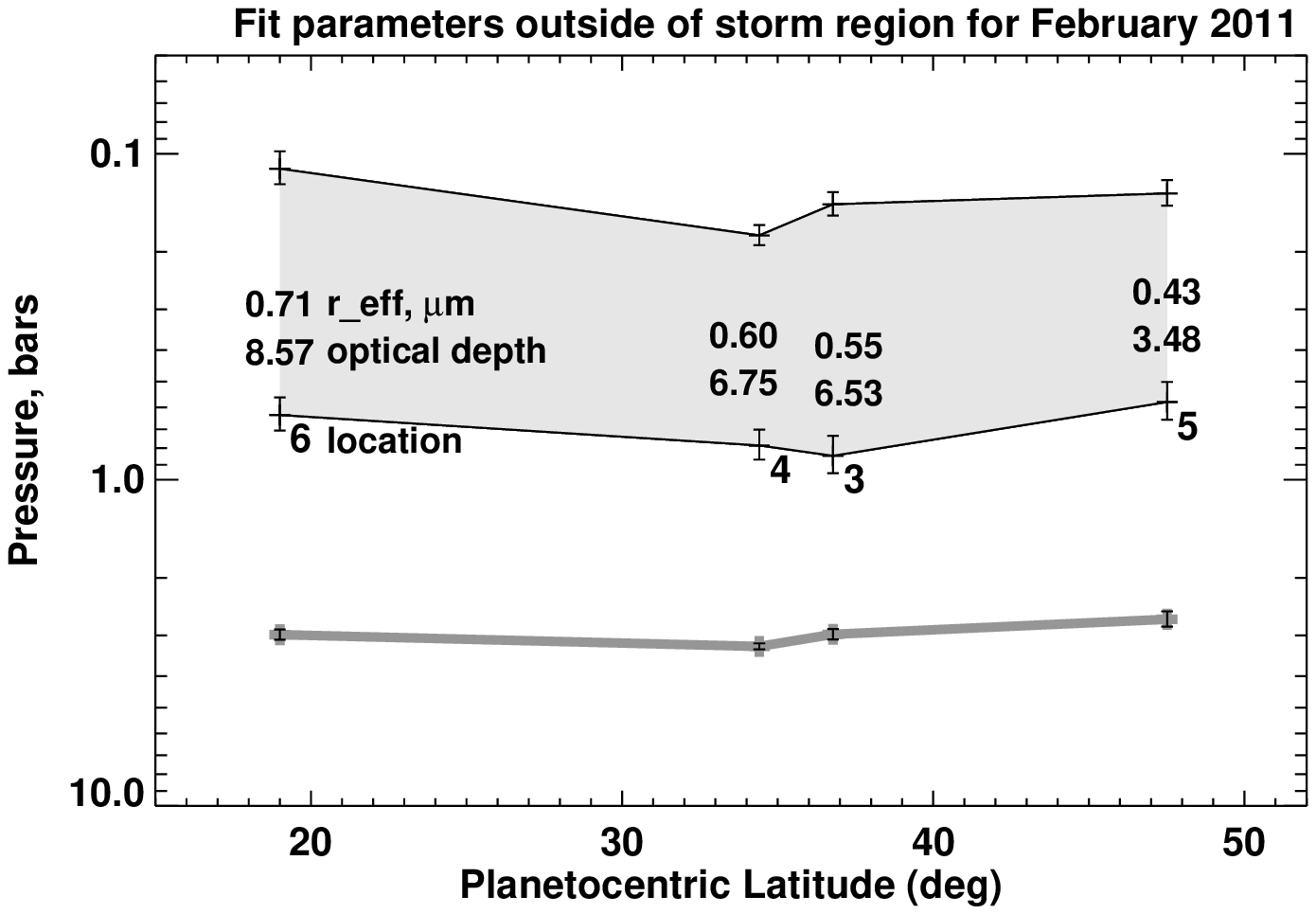}
\caption{Pressure boundaries of main cloud layer and effective pressure
of thermal barrier cloud vs latitude, with annotated values of location,
optical depth, and particle size. Locations refer to Fig.\ \ref{Fig:VIMSimages}.}
\label{Fig:parplot}
\end{figure}

Fits to spectra from all non-storm locations identified in
Fig. \ref{Fig:VIMSimages} and summarized in Table \ref{Tbl:parvals}
are graphically displayed in Fig. \ref{Fig:parplot}.  In these
non-storm fits, which sampled latitudes from 20\deg N to 48\deg N,
there is a trend for the main upper tropospheric haze layer towards
smaller particles and smaller optical depths with increased
latitude. In these fits there is no significant variation in the
effective height of the thermal blocking layer.  The significantly
worse fit to the location 6 spectrum, which is from the near
equatorial region, arises mostly from the low model I/F in the 1.5-1.6
\mum continuum region, which might be due to a small amount of
unaccounted-for absorption in the main cloud later in that broad
equatorial band.  Note that these fits exclude the latitude region
between 22\deg N and 30\deg N, where there has been a strong storm
influence, as indicated by the temporal changes evident in
Fig.\ \ref{Fig:ISSimages}.  Analyzing the character of that disturbed
region is left for future work.

\begin{table*}\centering
\caption{Cloud model parameter fits for February 2011 non-storm spectra.}
\vspace{0.15in}
\begin{tabular}{r c c c c }
\hline\\[-0.1in]
Parameter&  Location 3&  Location 4&  Location 5&  Location 6\\[0.05in]
\hline\\[-0.1in]
Planetocentric Lat., \deg & 36.78 & 34.41 & 47.52 & 19.00\\[0.05in]
Planetographic Lat., \deg & 42.57 & 40.10 & 53.32 & 22.94\\[0.05in]
East Long.,\deg & -116.09 & -115.89 & -110.63 & -102.90\\[0.05in]
Stratospheric haze $\tau\times 10^3$ & 8.6$\pm$0.9 & 8.6$\pm$0.8 & 7.5$\pm$0.7 & 15.0$\pm$1.8\\[0.05in] 
Main haze Ptop, mb & 143$\pm$11 & 178$\pm$12 & 132$\pm$11 & 111$\pm$12\\[0.05in] 
Main haze Pbottom, mb & 844$\pm$110 & 785$\pm$ 82 & 577$\pm$ 76 & 633$\pm$ 73\\[0.05in] 
Main haze $\tau$ & 6.53$\pm$0.81 & 6.75$\pm$0.66 & 3.48$\pm$0.36 & 8.57$\pm$0.83\\[0.05in] 
Main haze part. radius, \mum & 0.55$\pm$0.04 & 0.61$\pm$0.05 & 0.43$\pm$0.02 & 0.71$\pm$0.07\\[0.05in] 
Deep cloud Pbase, bar & 2.98$\pm$0.11 & 3.25$\pm$0.07 & 2.68$\pm$0.15 & 2.99$\pm$0.11\\[0.05in] 
$\chi^2$ & 204.08 & 149.87 & 169.30 & 430.19\\[0.05in] \hline
\end{tabular}
\parbox{5.in}{Uncertainties are 1-$\sigma$ values after adjustment
  for excess \chisq minimum, as described in text. \chisq values are
  for fits to the 48-wavelength sampled spectra. Locations refer to
  Fig.\ \ref{Fig:VIMSimages}.}
\label{Tbl:parvals}
\end{table*}

\subsection{Composition of the main cloud layer in non-storm regions}

None of the fits to the non-storm regions required any unusual
absorption in the vicinity of 3 \mumx, where \nhtx, \nthfx, and \nhfsh
have substantial absorption.  According to Fig.\ \ref{Fig:ratios},
there is a wide spread lack of 3-\mum particulate absorption in
regions not affected by the storm. It thus appears that
the main upper cloud layers outside the storm-affected regions are not composed
primarily of these materials.  This is a well known mystery that
remains to be definitively solved.  
It is quite amazing that the main cloud layer on Saturn that is
accessible to remote observations, which is of high optical depth and
thus represents a significant cloud mass, is apparently made of
materials that have no detectable near-IR spectral signatures and thus
do not seem to contain any significant contributions from the main
condensable gases on Saturn.

Explanations for how ammonia could really be the dominant condensable
in the main haze layer, even though it has not been detectable outside
of storm regions, have focused on three possible concealment
mechanisms: (1) large particle sizes, (2) non-spherical particle shape
effects, or (3) foreign coating or contamination.
\cite{West2009satbook} reviewed these explanations and rejected them
all, noting that a foreign contaminant would have to be the dominant
cloud mass to be effective at suppressing the \nht signatures. Our
calculations with composite particles are consistent with the
conclusions of West et al. and our inferred particle sizes are also
too small ($\sim$1\mum or less) to support (1).

The leading candidate for the upper tropospheric haze according to
\cite{Fouchet2009} is diphosphine (P$_2$H$_4$), which does not absorb
much at 3 \mumx, but has a distinctive double absorption peak
signature near 4.3 \mum and another at 6 \mum \citep{Nixon1956}. If
the 4.3-\mum peaks of diphosphine have roughly the same strength as
the double absorption peak of hydrazine (\nthfx), then that signature
should be visible even though it is partly obscured by \pht
absorption. However, though no such signature is seen either in or out
of storm regions, it is premature to completely rule out diphosphine
until refractive index measurements allow accurate radiative transfer
calculations to be made.  We can rule out hydrazine as the major
component of the main upper tropospheric haze because its distinctive
features would be easily detected, as apparent from the discussion in
the following sections.

\section{Modeling the 3-\mum absorption in storm cloud spectra.}

We first consider what compositions might be plausible, try simple
spectral fits using pure substances in a single main cloud layer, evaluate
their relative advantages, then consider more complex structures
including overlying sub-layers and composite particles.

\subsection{Optical properties of candidate cloud materials}

\begin{figure}[!htb]\centering
\includegraphics[width=3.35in]{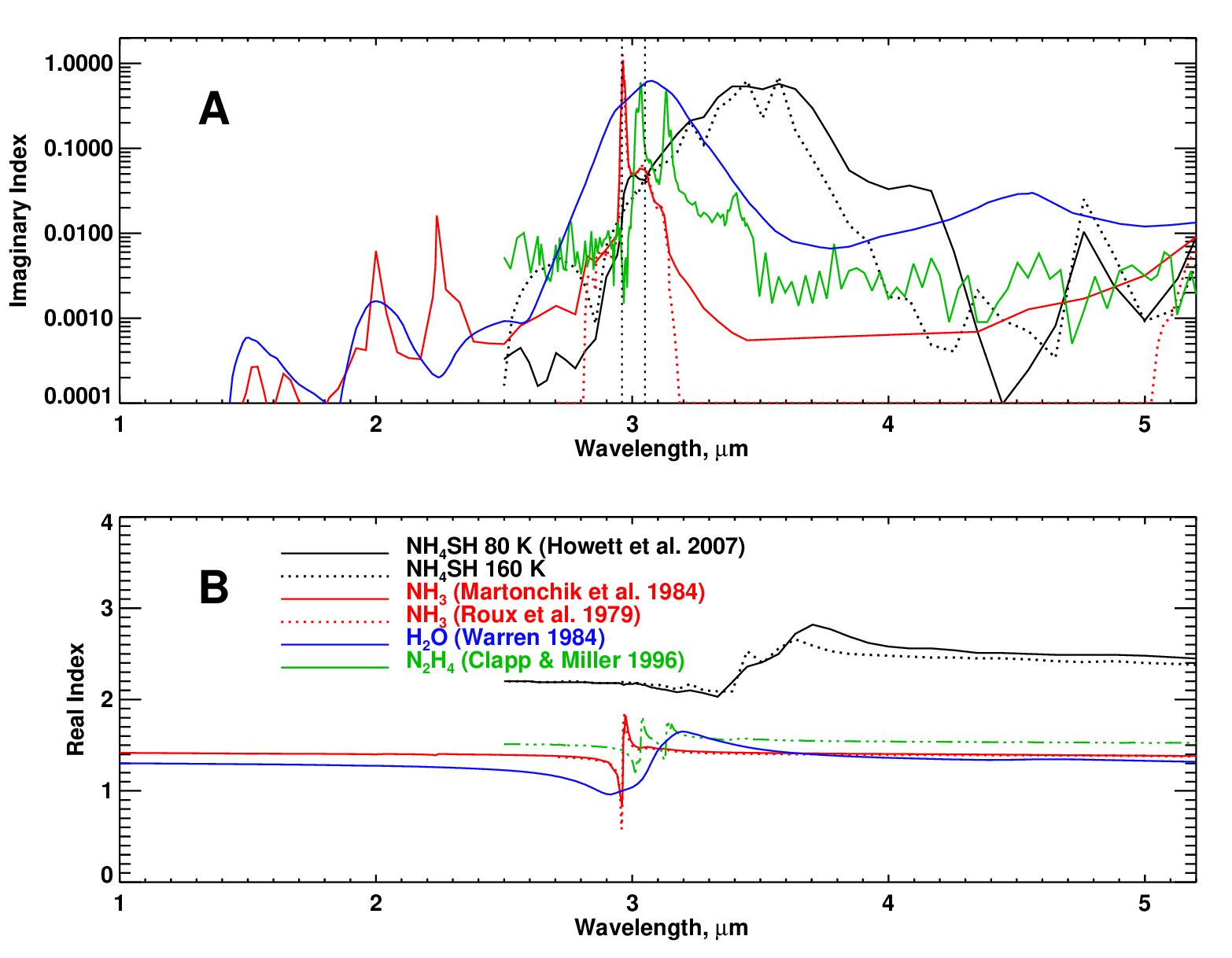}
\caption{Imaginary (A) and real (B) components of the refractive
index vs. wavelength for candidate 3-\mum absorbers, including H$_2$O
results of \cite{Warren1984}, \nht results of \cite{Roux1979} and
\cite{Martonchik1984}, NH$_4$SH results of \cite{Howett2007}, and
N$_2$H$_4$ results of \cite{Clapp1996}. Vertical dotted lines at 2.96
and 3.05 \mum indicate approximate locations of \nht absorption
features.}
\label{Fig:indexplot}
\end{figure}

Cloud materials expected in Saturn's atmosphere that also absorb light
in the 3-$\mu$m region, include NH$_3$ ice, NH$_4$SH (ammonium
hydrosulfide), water ice, N$_2$H$_4$ (hydrazine), and possibly solid P$_2$H$_4$ (diphosphine),
although it appears that the latter has only very weak bands in this
region \citep{Frankiss1968}.  Refractive index plots for these materials (except
diphosphine, which has not been characterized) are shown in Fig.\ \ref{Fig:indexplot}.
Among the more plausible storm particle components, NH$_3$ ice has the
strongest and sharpest absorption at 2 $\mu$m (the 2.25 $\mu$m
absorption feature would not be visible due to overlying gas
absorption).  Hydrazine, which is not expected to be very abundant on
Saturn, has a pair of strong absorption peaks between 3 and 3.2
$\mu$m, which would be very apparent unless cloud particles were very
large. Though diphosphine, which is expected to be more abundant on
Saturn \citep{Fouchet2009}, could plausibly also contribute to the
storm particle mix, we have no evidence in the VIMS storm spectra of
its distinctive double absorption peaks near 4.35 \mumx, and no
refractive index measurements that would allow us to include it in
quantitative radiative transfer models.
\nhfsh has an absorption that is roughly comparable to that of NH$_3$
at 3 $\mu$m, but lacks a sharp spectral feature and continues to
increase beyond 3.1 $\mu$m while NH$_3$ absorption drops
significantly.

\begin{figure*}[!htb]\centering
\includegraphics[width=3.2in]{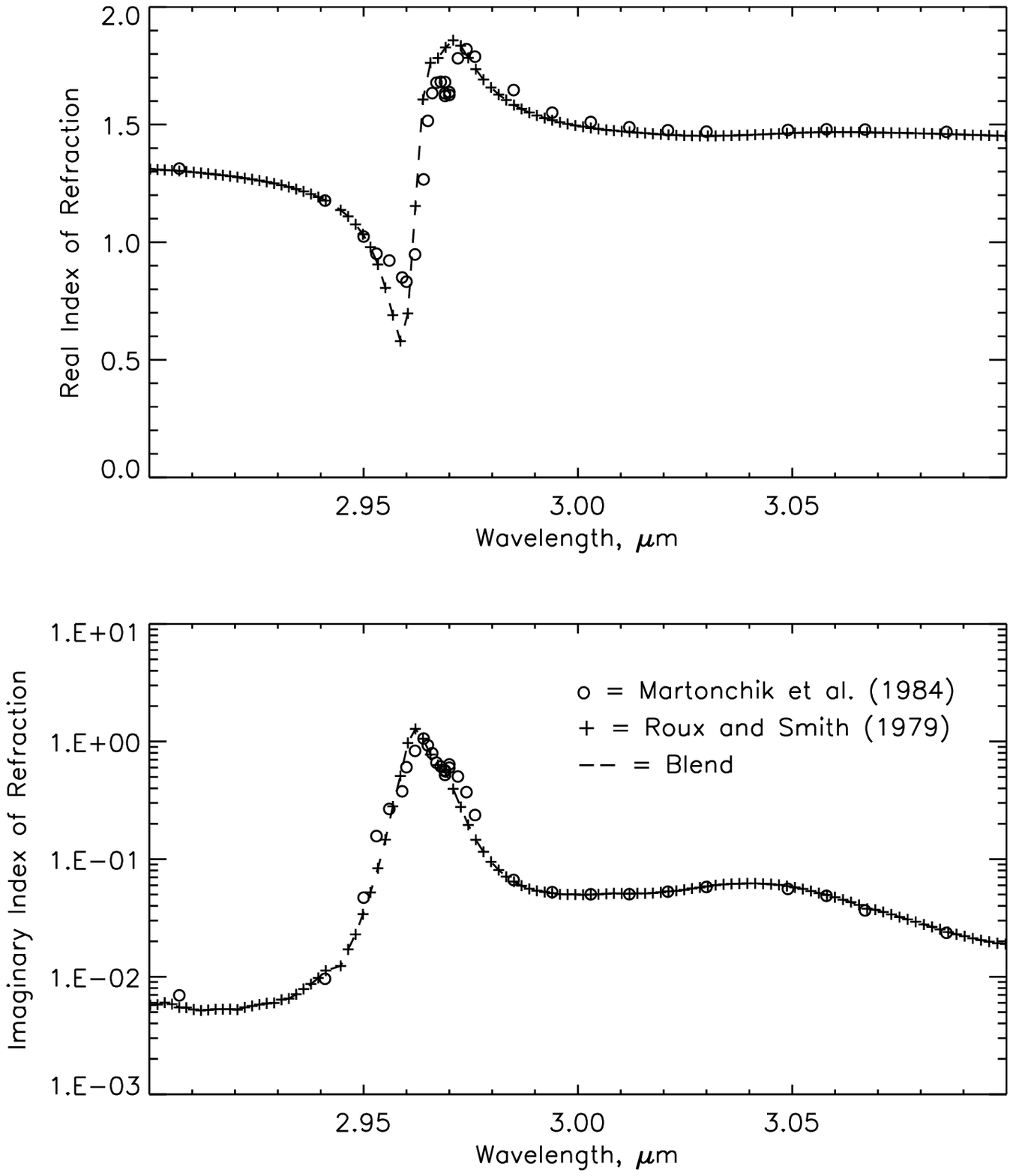}
\includegraphics[width=3.2in]{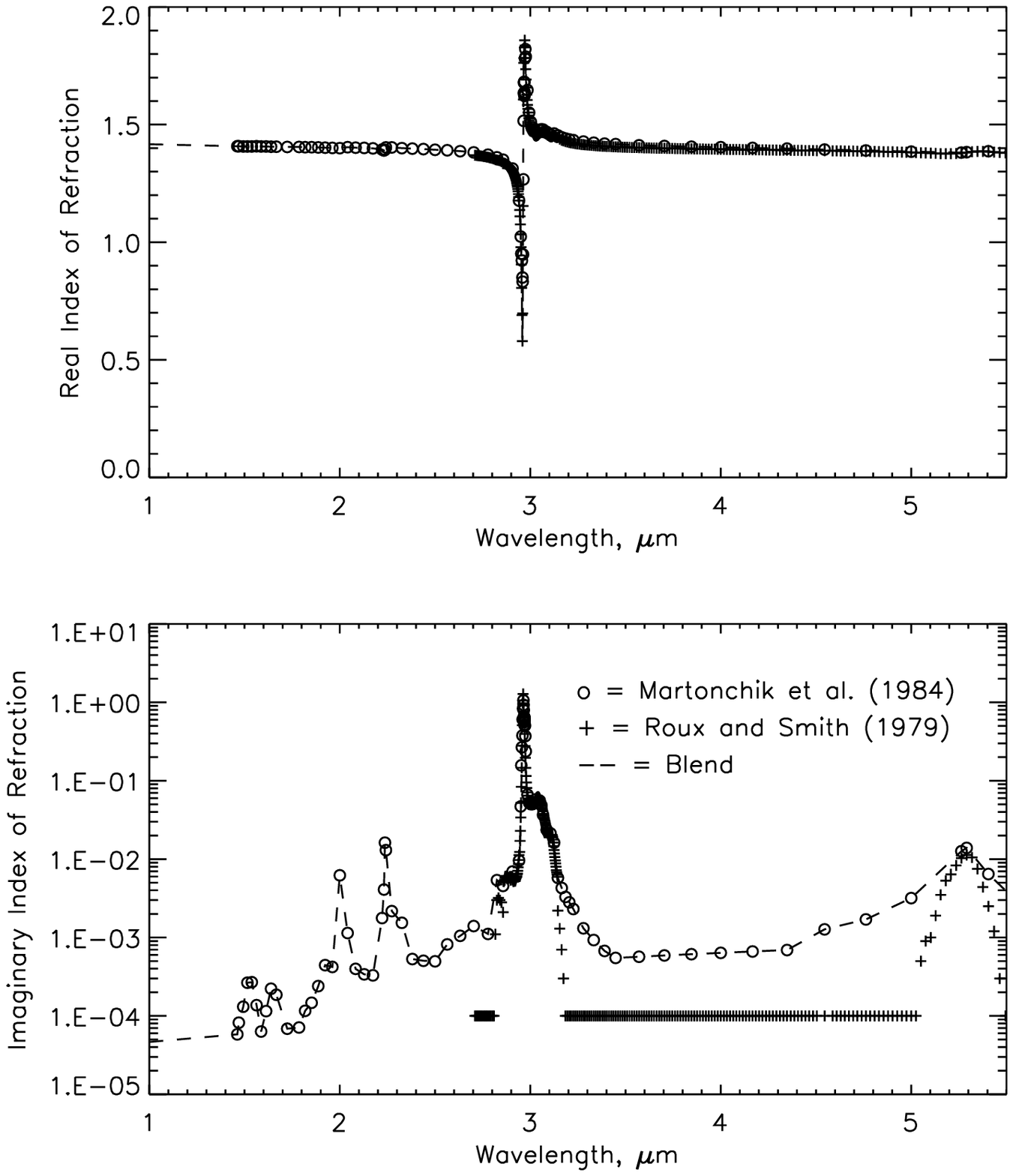}
\caption{Comparison of \cite{Martonchik1984} and \cite{Roux1979} measurements of
the refractive index of \nhtx. Our second blended version (dashed curve), uses the former
everywhere except between 2.88 \mum and 3.14 \mumx, where the latter is used. \cite{Roux1979}
provides a better match to VIMS observations near 2.9 \mumx, while \cite{Martonchik1984} provides
a better match near 5 \mumx. }
\label{Fig:nh3index}
\end{figure*}

There is some disagreement about the detailed refractive index
variation with wavelength for \nhtx. A comparison of results from
\cite{Martonchik1984} and from \cite{Roux1979} is provided in
Fig.\ \ref{Fig:nh3index}.  Because \cite{Roux1979} provides better
spectral resolution in the 2.7-3.1 \mum region and provides no
information below 2.8 \mumx, we initially used \cite{Martonchik1984}
for $\lambda < 2.8$ \mum and \cite{Roux1979} for longer wavelengths.
These two references also disagree on the level of absorption near 5
\mumx.  We later used a different blend in which we used
\cite{Martonchik1984} everywhere except between 2.88 \mum and 3.14
\mumx, where \cite{Roux1979} is used.  This provides improved fit
quality for some models. There is also evidence for small variations
of optical properties with temperature \citep{Sill1980}, which we
cannot account for and thus are forced to ignore.

\subsection{Initial spectral fits with pure particles}

The five substances we considered in preliminary calculations with
single-composition particles are these: (1) conservative substance
with refractive index n=1.4+0i, (2) \nhtx, (3) \nhfshx, (4) \nthfx,
and (5) water ice.  As indicated in Fig.\ \ref{Fig:indexplot},
substances 2-5 all have significant absorption in the 3-\mum region.
The question we are trying to answer is which substance is dominant,
if any, and is there evidence of contributions from more than one
substance?  The great depth of vertical convection suggested by
numerical models of the dynamics of this feature \citep{Hueso2004Icar}
makes it plausible to consider that even water ice might reach levels
of the main cloud layer.  Intense lightning strikes inferred from
static electric discharge (SED) events measured by the Cassini Radio
and Plasma Wave instrument (RPWS) occurred mainly near the head of the
storm and indicate that this is a deep convective storm area energized
by latent heat of water condensation in the 10-20 bar level. Other
lightning strikes seen optically by the ISS instrument confirm their
vertical location 125-250 km below the upper cloud layer
\citep{Dyudina2010GRL}.

\begin{figure*}[!htb]\centering
\hspace{-0.05in}\includegraphics[width=6.2in]{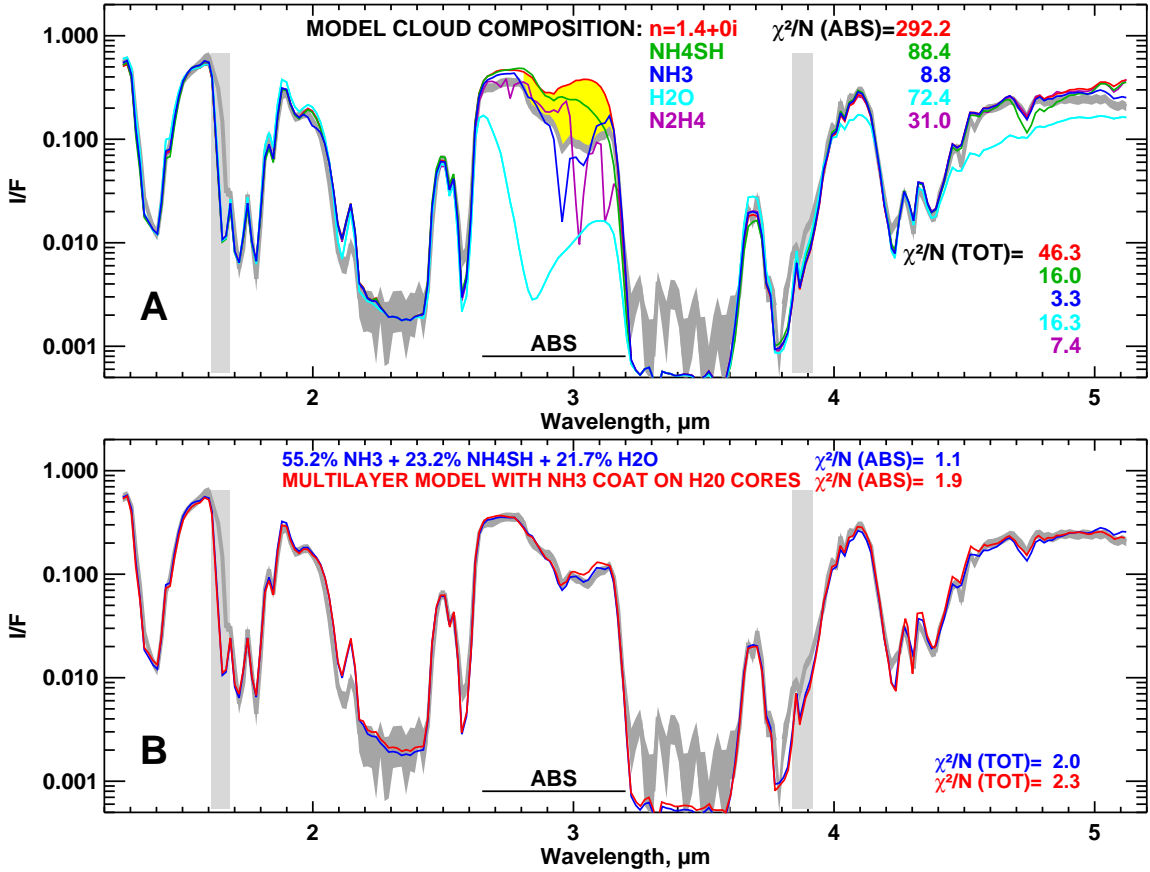}
\caption{({\bf A}) A storm head spectrum from (location 1 in
  Fig. \ref{Fig:VIMSimages}), and fits to that spectrum using a main
  cloud layer consisting of either conservative particles (red), pure
  \nhfsh particles (green), pure \nht particles (blue), or pure \hto
  particles (cyan).  The vertical gray bars indicate regions excluded
  from our analysis because order sorting filter joints lead to
  uncertain responsivity corrections at those wavelengths.  Relative
  fit quality for each entire pure spectrum (excluding the two regions
  indicated by vertical gray bars) is indicated by the $\chi^2$/N
  (TOT) values, and for the 2.65 \mum - 3.2 \mum region of particulate
  absorption by $\chi^2$/N (ABS) values, where there are 33 points in
  the ABS region and 223 in total.  The specific fit parameters for
  each spectrum are given in Table\ \ref{Tbl:purepars}. The yellow
  region between the conservative particle spectrum (red) and the VIMS
  measured spectrum is due to cloud particle absorption.  ({\bf B}) As
  in A except that the blue model spectrum is a linear combination of
  pure spectra, with fractions adjusted to minimize \chisq in the
  particle absorbing (ABS) region, and the red spectrum is for Model 3
  (discussed in Section \ref{Sec:Model3}), which is a multi-layer
  horizontally homogeneous model with the main upper
  tropospheric haze layer containing an upper sub layer of
  conservative particles and a lower sub layer of ammonia-coated water
  ice particles, as defined in Table \ref{Tbl:extendedh2o}.}
\label{Fig:pure}
\end{figure*}

To gain a better understanding of how various plausible particle
constituents might contribute to the 3-\mum absorption feature, we
first try to fit the storm spectra using pure substances in a single
middle cloud layer, leaving for later the issue of possible composite
particles and/or multiple layers of different composition.  We assumed
that the middle layer was composed of pure spherical particles, then
adjusted the optical depth, particle size, and top and bottom
pressures to optimize fits for the region from 1.268 \mum to 5.15
\mumx, but ignoring the region from 2.65 \mum to 3.2 \mumx, where
particulate absorption is most prominent, as well as ignoring the two
narrow regions indicated in Fig. \ref{Fig:pure}, where VIMS measurements have large
uncertainties (described in Section \ref{Sec:artifact}). That allows the vertical structure
of the aerosols to be primarily constrained by methane gas absorption
and not much affected by particulate absorption.  We then compared the
model spectra with measured spectra in the 2.65 \mum - 3.2 \mum region
to determine which provided the best compositional match.

The fits for the five pure substances are displayed in
Fig. \ref{Fig:pure}A.  The fit parameters and fit
quality are summarized in Tables\ \ref{Tbl:purepars} and
\ref{Tbl:purequal}. We evaluate the overall fit using $\chi^2$
computed  as described for the non-storm regions. 
he conservative particles with n=1.4 (Fig.\ \ref{Fig:pure}A) provide
a reasonably good spectral fit at all wavelengths except in the 3-\mum
region, where the model has a small atmospheric absorption due to
phosphine and the measurements require a much larger absorption over a
broader range that must be due to cloud particles.  The phosphine
absorption is particularly small here because the cloud layer
containing these particles is relatively high (140-550 mb) and
optically thick ($\tau$=9 at 2 \mumx).  The residual absorption
(colored yellow in Fig. \ref{Fig:pure}A), which is the difference between the
conservative model (red curve) and the measured VIMS spectrum (thick
gray curve in Fig.\ \ref{Fig:pure}), thus must be provided predominantly by
particle absorption, and entirely so between 3.05 and 3.1 \mumx.  This
also shows that the particle absorption is significant.  

The ammonia spectrum provides by far the best match to the measured
spectrum of the storm head, yielding a \chisq value in the absorbing
region that is 3.5 times better (smaller) than that of the next best fit.  While
ammonia particles provide the best fit among these simplistic
single-component models, the resulting model spectrum still contains
substantial differences from the VIMS measured spectrum, especially
the sharp ammonia ice absorption feature near 2.96 \mumx, but also the
excess brightness near 3.1 \mumx. The model spectrum also has a small
dip at 2 \mumx, where \nht has another sharp, but much weaker,
absorption feature, while a corresponding dip is not seen in the
measured spectrum.

\begin{table*}\centering
\caption{Adjustable fit parameters for different main cloud compositions.}
\vspace{0.2in}
\begin{tabular}{c c c c c c c c}
\hline\\[-0.1in]
               & \multicolumn{4}{c}{Main cloud layer parameter values}   \\
Main cloud     & Top  & Bottom & Particle & 2-\mum Optical  & \\
 composition & P, mb & P, mb   & radius, \mum   & depth (2-\mumx) & \chisq  \\
\hline\\[-0.1in]
  n=1.4+0i           & 129$\pm$3 & 647$\pm$40 & 1.34$\pm$0.07 & 12.7$\pm$0.7 & 538  \\ 
  \nht               & 133$\pm$3 & 849$\pm$90 & 1.35$\pm$0.06 & 18.7$\pm$2.1 & 426 \\ 
  \nhfsh             & 136$\pm$3 & 945$\pm$120 & 0.61$\pm$0.01 & 7.5$\pm$0.7 & 591 \\ 
  H$_2$O             & 182$\pm$5 & 401$\pm$60 & 0.83$\pm$0.10 & 16.4$\pm$3.7 & 1274\\ 
  \nthf             & 138$\pm$3 & 826$\pm$84 & 1.13$\pm$0.04 & 13.8$\pm$1.5 & 479\\ 
\hline\\[-0.1in]
\end{tabular}
\parbox{4.4in}{Fixed parameters included the pressure (2 mb), and
  2-\mum optical depth (0.01) of the stratospheric haze (assumed to be
  a conservative, isotropic scatterer with $\lambda^{-3.5}$ wavelength
  dependence), and the pressure (1.7 bars), optical depth (100),
  single-scattering albedo (0.9), and asymmetry parameter (0.6) of the
  deep cloud.  Mie scattering was assumed for the main cloud layer,
  with equal gas and particle scale heights. Uncertainty limits given
  here are formal errors useful for relative comparisons only, with no
  correction to account for excess \chisq minima. \chisq values listed
  here are for fits to the 48-wavelength sampled spectra. }
\label{Tbl:purepars}
\end{table*}

\begin{table*}\centering
\caption{Fit quality for single-particle spectra and best fit combination spectra.}
\vspace{0.15in}
\begin{tabular}{|c c c c| c c| c |}
\hline\\[-0.1in]
\multicolumn{4}{|c|}{Component fractions}&\multicolumn{2}{|c|}{Fit quality ($\chi^2/N$)}&\\
n=1.4+0i & \nht & \nhfsh & H$_2$O & 1.26-5.16 \mum &  2.65-3.1 \mum & Notes\\
\hline\\[-0.1in]
   1     &      &      &       &   46.3       &   292.2   &1      \\
        &   1   &      &       &   3.3       &   8.8     &1      \\
        &      &    1  &       &   16.0      &  88.4     &1      \\
        &      &      &   1    &   16.3       &   72.4    &1      \\
  &  0.55$\pm$0.04 &  0.23$\pm$0.02 & 0.22$\pm$0.02   &   2.0 &   1.1  &2   \\
  0.13$\pm$0.02   &  0.62$\pm$0.03 & & 0.35$\pm$0.03   &  2.4  &  2.9  &2   \\
  0.00$\pm$0.00   &  0.83$\pm$0.02 &  0.17$\pm$0.02  & &  3.7 &   5.2  &2   \\
\hline
\end{tabular}
\label{Tbl:purequal}
\parbox{4.5in}{$^1$Fits for single particle types in the main cloud
  layer, with parameters given in
  Table\ \ref{Tbl:purepars}.\\ $^2$These are linear combination fits
  with blank columns indicating particle types that were not included
  in the fit. Here \chisq values were computed using the simplified
  error model described in the text.  Note the significant degradation
  in fit quality when water ice is excluded.}
\end{table*}

Turning to the option of pure \nhfsh particles (green curve in
Fig.\ \ref{Fig:pure}A), we see that the overall fit is better than
that obtained with conservative particles, mainly due to the better
fit in the 3-\mum region, but the fit quality in the 2.65-3.2 \mum
region (\chisq/N (ABS) = 88.4) is ten times worse than obtained from
pure \nhtx.  It does have a potentially useful characteristic,
however.  Its I/F at the edge of that window ($\approx$3.15 \mum) is
lower than obtained from \nhtx, so that combining the two might
improve the overall fit quality. While hydrazine (\nthfx) provides a
decent overall spectral fit, its double absorption peaks in the 3-3.2
\mum region are a poor match to the observed absorption and
also provide little help as a minor contributor.

The fit for pure water ice (cyan curve in Fig.\ \ref{Fig:pure}A) shows that water is a very strong
absorber, with a spectral signature completely different from that of the other substances,
and especially different from what is needed to match the observed spectrum.  However,
it does turn out to have useful properties when we consider composite fits.  We often
find that good fits using \nht and \nhfsh result in the window regions outside of
the 2.65-3 \mum window being relatively too dark.  By adding a small amount of
the water spectrum, we can brighten these window regions without brightening
the 3-\mum region because of water's strong absorption at those wavelengths.
Incidentally, it may appear strange that water ice fits the observed
spectrum better than \nhfsh according to \chisq values, yet visually appears
to be a much worse fit.  The visual appearance is misleading because the plots
use a logarithmic y axis, while \chisq calculations use a linear weighting of
differences (squared). 

\subsection{Linear combination spectral fits}

We next consider linear combinations of pure spectra to see if mixed
particle compositions could provide a better match than pure particles.
These combinations can be thought of as spatially heterogeneous
mixtures, perhaps as different particles visible in spaces between
clouds of other particles. They are also suggestive of what might
happen if an optically thin layer of one cloud type would overlie a
layer of another cloud type (but not vertically separated by large
distances), or if a cloud consisted of composite particles, perhaps with
a core of one (or more) materials with a condensed shell of another
material. 
The best linear combination, shown by the blue curve in
Fig.\ \ref{Fig:pure}B, uses a mix of 55$\pm$4\% \nhtx, 22$\pm$2\%
water ice and 23$\pm$2\% \nhfshx, where percentages represent
fractional area coverage.  This improves the fit in the absorbing
region to $\chi^2/N (ABS) =1.1$, a factor of 8 better than provided by
pure \nht (see Fig.\ \ref{Fig:pure} caption).  A big improvement is
also obtained using the conservative (n=1.4) material in place of
\nhfshx, with a mix of 62$\pm$3\% \nhtx, 35$\pm$3\% water ice, and
13$\pm$2\% n=1.4, although the improvement is only by a factor of five
in this case.  (Uncertainties quoted here are deviations that produce
$\Delta$\chisq=$\chi^2_{MIN}/N$, where $N$ is the number of degrees of
freedom.) The best linear combination without water ice produced
$\chi^2/N (ABS) = 5.2$, which is a factor of 5 worse than produced by
the best combination with water ice.  The water ice component reduces
the excess I/F gradient between 3 \mum and 3.15 \mum and the excess
I/F otherwise present between 2.65 \mum and 2.85 \mumx.  The third
component (either n=1.4 or \nhfshx) serves to partially fill in the
deep minimum produced by pure \nht at 2.96 \mumx.  The linear
combination analysis clearly favors \nhtx, water ice, and \nhfsh in
that order. Our next step was to determine whether what we think is a
somewhat more realistic physical configuration, i.e. a horizontally
homogeneous cloud structure, could achieve similar fit quality.  The
red curve in Fig.\ \ref{Fig:pure}B is for the last of three such
structures we tried.

\subsection{Models with composite particles}

Numerical experiments with thick layers of coated particles containing
substantial fractions of \nht as the main tropospheric haze layer
produced excessively strong \nht signatures unless the fraction of
\nht was so small that the desirable weaker features were also
eliminated.  We found a more successful strategy was to split the main
tropospheric haze layer into two parts, one basically conservative,
and one dominated by \nhtx. There are two relatively plausible
implementations of this approach that are both capable of suppressing
the stronger \nht features while retaining enough of the \nht
character to match the VIMS observations.  The first (Model 1) is to
put an optically thin layer of \nht particles above a layer with
relatively flat spectral properties in the 3-\mum region.  The second
(Model 2) is to put an optically thin layer of conservative particles
above an optically thick layer of particles containing a significant
component of \nhtx.  In the following we consider both options and
third (Model 3) that is a variant of the second.  The model spectra
are compared to the measured storm spectrum in
Figs.\ \ref{Fig:waterfit}-\ref{Fig:extendedh2o}, with model vertical
structures displayed in Fig.\ \ref{Fig:structurecomp}. The complexity
of these models, involving different materials, different radii,
different core fractions and combinations of core and shell materials,
as well as optical depth and pressure of each layer, is so large that
we cannot demonstrate uniqueness for any of these solutions.

\subsubsection{Model 1: an \nht layer on top.}

\begin{figure*}[!htb]\centering
\includegraphics[width=6.2in]{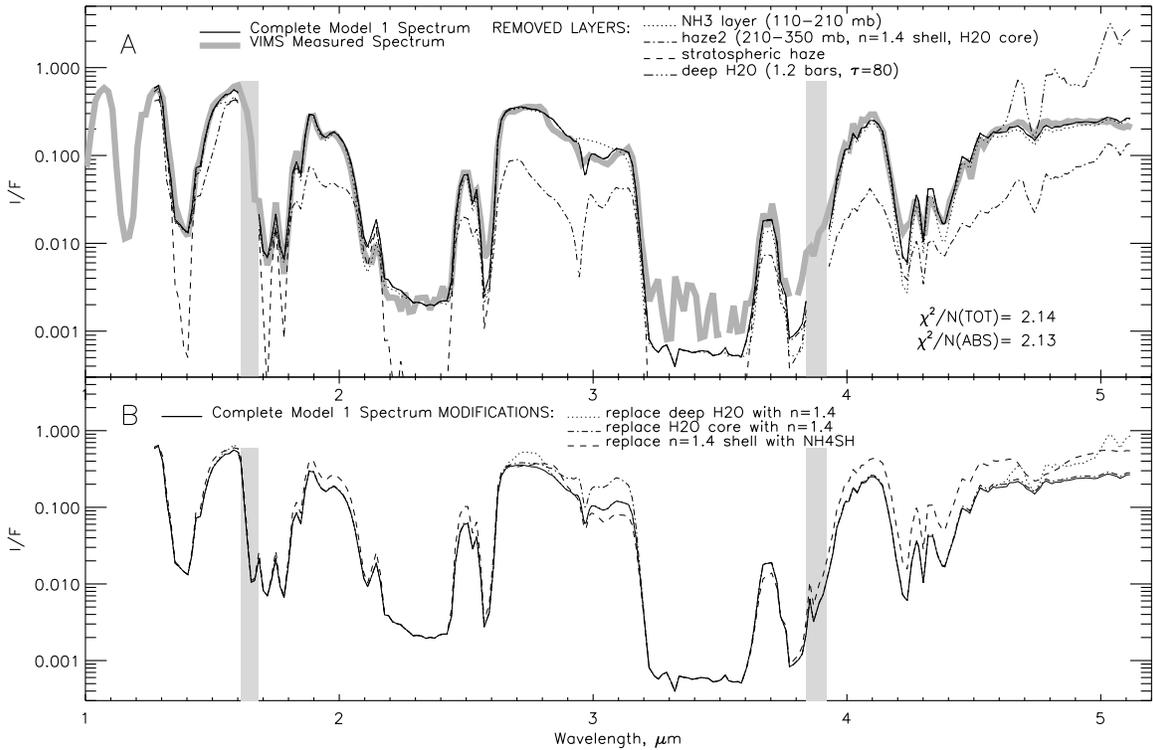}
\caption{A: VIMS measured spectrum of the storm head (thick gray line) in comparison with
a complete model spectrum (solid line), and model spectra with different layers
removed one at a time (see text for discussion). B: Model spectrum (solid curve)
compared to spectra for modified models in which the deep water cloud is replaced
by a cloud with n=1.4 (dotted), or the H$_2$O core of the 210-350 mb haze is
replaced by an n=1.4 core (dot-dashed), or the n=1.4 shell of the 210-350 mb
haze particles is replaced by an \nhfsh shell (dashed). See text for discussion.}
\label{Fig:waterfit}
\end{figure*}

The first model we considered consists of a deep water cloud at 1.2
bars, a main haze sublayer (haze2 in Fig.\ \ref{Fig:waterfit} legend)
of composite particles extending between 210 mb and 350 mb, an \nht
haze layer from 110 mb to 210 mb, and a stratospheric haze that is of
little physical significance because of uncertain I/F measurements at
low I/F values.  The parameters of this model are defined in
Table\ \ref{Tbl:waterparams}.  In Fig.\ \ref{Fig:waterfit} the
corresponding model spectrum is compared to the spectrum from the head
of the storm. In Fig.\ \ref{Fig:waterfit}A we see that the model
spectrum does a good job of capturing the main features of the
measured spectrum, including the spectral features between 2.9 and 3.1
\mumx, which are due primarily to \nht ice.  No other plausible
material in Saturn's atmosphere can produce the sharp feature near
2.96 \mum or the secondary bump at 2.05 \mumx.  The \nht layer has a
relatively small optical depth ($\tau$=1) so that the underlying
bright layer can limit the depth of the sharp absorption feature at
2.96 \mumx.  These \nht ice particles are assumed to be spheres of
radius 1.5 \mumx.  We found that significantly larger particles tended to
produce to large an I/F at longer wavelengths, while smaller particles
made it more difficult to match spectral features near 3 \mumx.

The underlying layer that provides most of the reflected light visible
outside the atmosphere is here composed of composite particles.  After
trying fits with spectrally flat particles in this layer, we were
unable to keep the I/F at 3.1 \mum low enough without causing problems
at other wavelengths.  An example is shown in
Fig.\ \ref{Fig:waterfit}B, which shows the effects of several
different modifications to the nominal compositional model.  In one
case we used a homogeneous particle of refractive index n=1.4
(computed as a composite particle in which both core and shell are the
same), which produces the spectrum shown by the dot-dash curve.  This
is seen to be too bright at 2.9-3.15 \mumx, while the fit over the
rest of the spectrum is very good.  The nominal model, which uses a
small core of H$_2$O (40$\pm$5\% of the total radius), takes care of
this problem nicely.  We also tried to use \nhfsh as a shell material
(the dashed curve in Fig.\ \ref{Fig:waterfit}B), but this causes
problems at many places in the spectrum.  Although with some minor
adjustment it could produce a reasonable fit in the 2.9-3.1 \mum
region, it tends to be too bright at 1.9-2.0 \mumx, at 2.85 \mumx, and
beyond 3.9 \mumx, as well as being too dark near 3.7 \mum because of
an absorption by \nhfsh in this region.  Some of these problems can be
fixed by adjusting other parameters, but we could not find a fix for
all of them simultaneously.  It is much easier to fit the spectrum
using a shell composed of the n=1.4 material that seems to be the main
component of clouds in other (non-storm) regions of Saturn's
atmosphere.

We also considered a different composition for the deep and optically
thick cloud that is mainly needed to block thermal radiation beyond
4.8 \mumx.  Using the n=1.4 material for that cloud produces too high
an I/F in the 2.7-2.8 \mum region as well as allowing too much thermal
radiation to leak through near 5 \mumx.  To fix the thermal leak would
require either greatly increased optical depth or an additional
absorption.  Water fixes both these problems because it has absorption
in both regions of the spectrum, while still providing the needed high
reflectivity for the shorter wavelength windows at 1.26 \mum and 1.62
\mumx.

\begin{table*}\centering
\caption{Fit parameters for composite storm cloud structure Model 1.}
\vspace{0.15in}
\begin{tabular}{c c c c c }
\hline\\[-0.1in]
Particle     & \multicolumn{2}{c}{Layer boundaries} & Particle   \\
 Composition & Ptop   & Pbottom   & radius      & Optical depth   \\
\hline\\[-0.1in]
  \nht        & 110 mb & 210 mb & 1.50 \mum & 1.0  \\[0.1in]
\parbox{0.85in}{n=1.4+0i shell on \hto core}     & 210 mb & 350 mb & 
         \parbox{0.55in}{1.5 \mum with\newline 0.6 \mum core} & 8.0 \\[0.15in]
  \hto      & 1000 mb & 1200 mb & 2.0 \mum & 80 \\[0.1in]
\hline\\[-0.1in]
\end{tabular}
\parbox{4.75in}{We also included a stratospheric haze between 0.5 mb
  and 2 mb, which here was treated as a double-HG scatterer with
  $\tau$=0.01 at 2 \mumx, and a $\lambda^{-3.5}$ wavelength dependence.  Nominal
  gas mixing ratios and profiles were assumed as described elsewhere,
  except that the \pht scale height was assumed to be 50\% of the gas
  scale height.}
\label{Tbl:waterparams}
\end{table*}

\subsubsection{Model 2: conservative particles on top.}

\begin{figure*}[!htb]\centering
\includegraphics[width=6.2in]{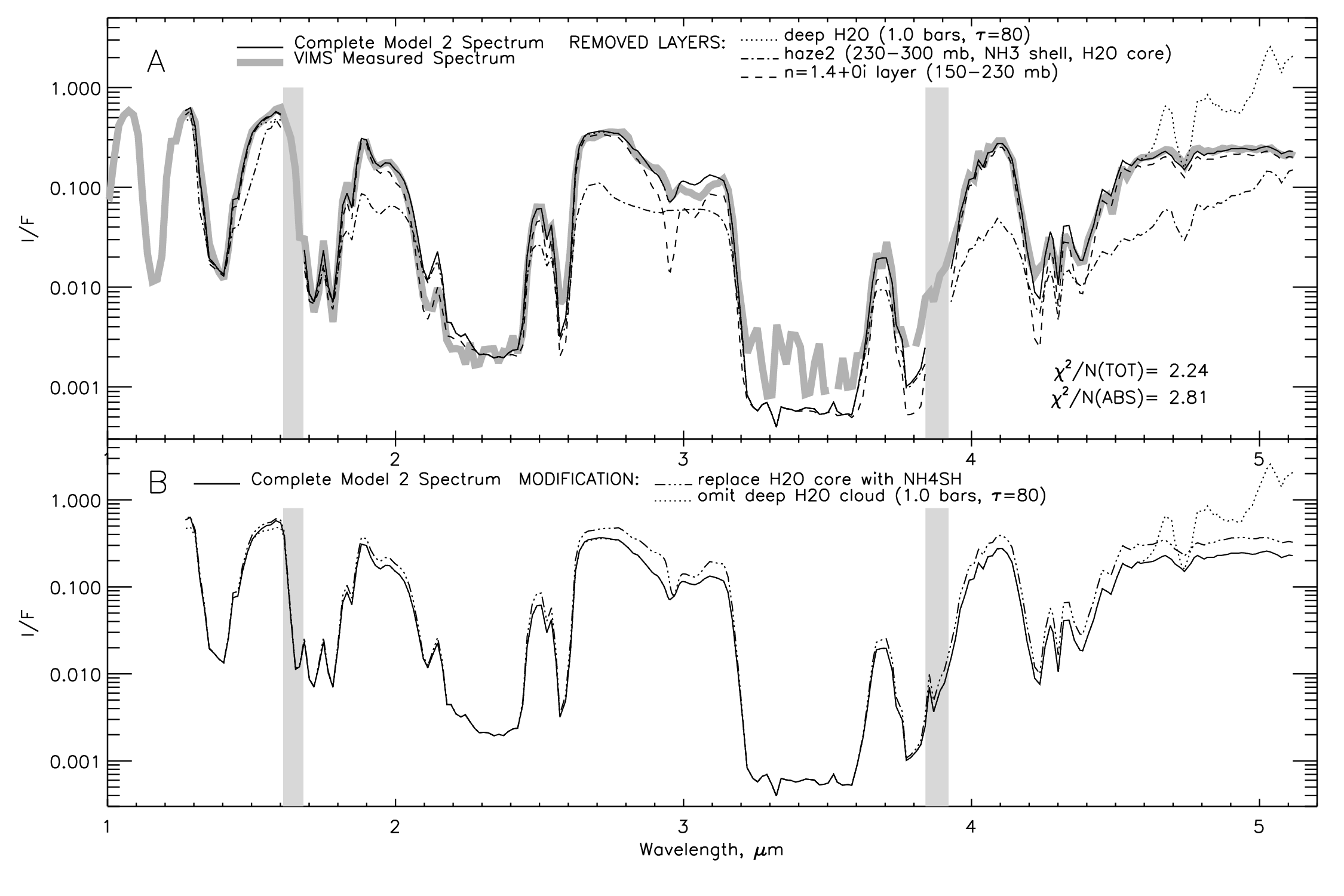}
\caption{A: VIMS measured spectrum of the storm head (thick gray line)
  in comparison with a complete model spectrum (solid line) for the
  alternate composite model defined in Table\ \ref{Tbl:waterparams2};
  model spectra with different layers removed one at a time are also
  shown (see text for discussion). The removal of the stratospheric
  haze layer (not shown) has essentially the same effect as shown in
  Fig.\ \ref{Fig:waterfit}. B: Alternate model spectrum (solid curve)
  compared to spectra for modifications in which the deep water cloud
  is removed (dotted), or the H$_2$O core of the 230-300 mb haze is
  replaced by an \nhfsh core (triple dot-dashed). See text for
  discussion.}
\label{Fig:waterfit2}
\end{figure*}

Perhaps a somewhat more realistic physical model is the alternate
structure summarized in Table\ \ref{Tbl:waterparams2}, with the
corresponding spectrum displayed in Fig.\ \ref{Fig:waterfit2}.  In
this case the top part of the main cloud layer consists of a pure
conservative substance, which we assume to be the same material seen
outside of storm regions and is assumed to have a refractive index of
n=1.4+0i.  The \nht appears in the lower part of the main cloud layer
as a coating on a small core of \hto (also about 40\% of the total
radius).  The deep layer is composed of pure \hto ice, as in the
previous composite model.  This produces a spectral match which is
comparable to the first model, but is more easily understood
physically.  If the storm convection begins at the water cloud level,
then as the water particles rise and freeze they will eventually serve
as (large) condensation nuclei for \nht which will eventually reach
saturation and form a coating on the water cores.  This makes more
sense than forming coatings of n=1.4+0i material, since that material
is not even known to be formed by condensation at the relevant level
of the atmosphere. The top layer of n=1.4+0i material can perhaps be
understood as part of the original main cloud layer pushed to higher
altitudes by the convecting tower that brings up water from the deep
levels of the atmosphere.  This layer has the important spectral
effect of filling in the deep minimum at 2.96 \mum that would
otherwise be visible due to the layer below it, as can be seen from
the spectrum in Fig.\ \ref{Fig:waterfit2}A that is obtained when the
layer is removed. Although we do not seem to need \nhfsh or find it
useful either in the form of a pure particle or as a coating or core
in a two-component composite particle, it is expected that particles
are likely to contain a water core, a layer of \nhfsh, followed by a
layer of \nhtx.  Unfortunately, we do not have tools for modeling such
particles.

\begin{table}\centering
\caption{Fit parameters for storm cloud composite structure Model 2.}
\vspace{0.15in}
\begin{tabular}{c c c c c }
\hline\\[-0.1in]
Particle     & \multicolumn{2}{c}{Layer boundaries} & Particle &Optical  \\
 Composition & Ptop   & Pbottom   & radius      & depth   \\
\hline\\[-0.1in]
  n=1.4+0i        & 150 mb & 230 mb & 1.50 \mum & 1.1  \\[0.1in]
\parbox{0.75in}{\nht shell on\newline \hto core}     & 230 mb & 300 mb & 
         \parbox{0.75in}{1.5 \mum with 0.6 \mum core} & 9.0 \\[0.15in]
  \hto      & 900 mb & 1000 mb & 2.0 \mum & 80 \\[0.1in]
\hline\\[-0.1in]
\end{tabular}
\parbox{4.25in}{Stratospheric haze and \pht scale height as in Table\ \ref{Tbl:waterparams}}.
\label{Tbl:waterparams2}
\end{table}

\subsubsection{Model 3: conservative top layer with an extended deep cloud.}\label{Sec:Model3}

\begin{figure*}[!htb]\centering
\includegraphics[width=6.2in]{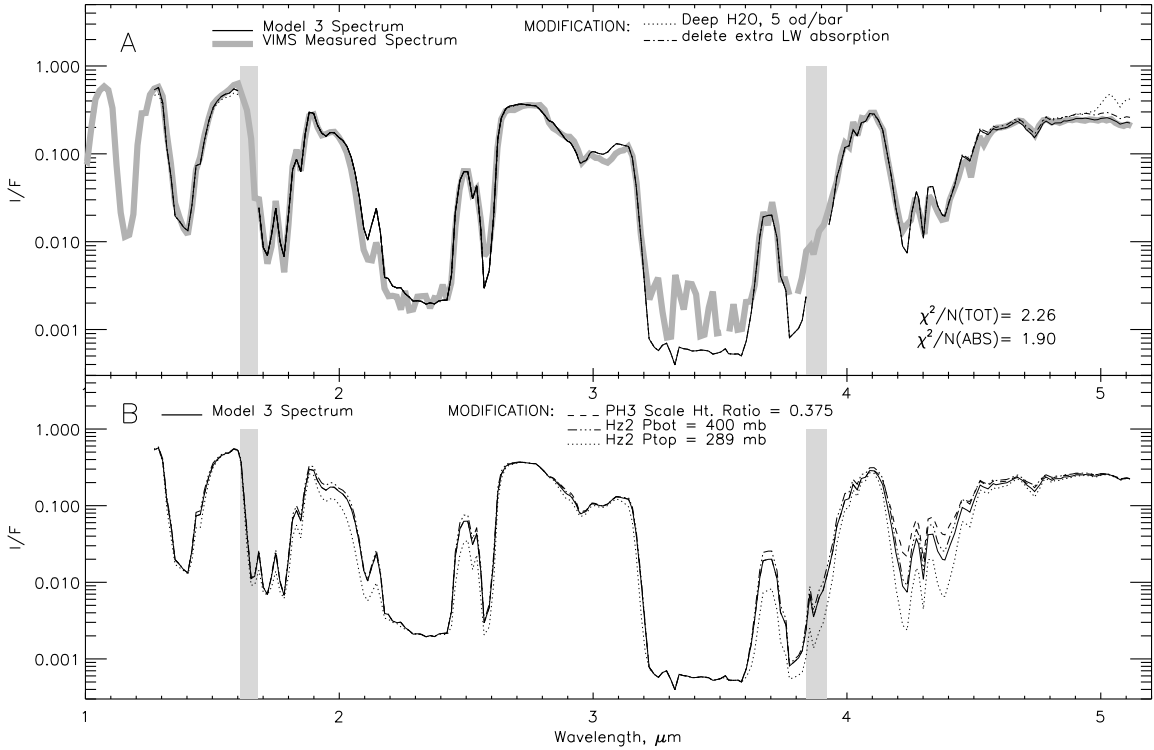}
\caption{A: VIMS measured spectrum of the storm head (thick gray line)
  in comparison with the model 3 spectrum (solid line), and modified
  spectra with different values for the optical depth per bar of the
  vertically extended deep water cloud.  B: Model 3 spectrum (solid
  curve) compared to spectra for modified models with slightly
  different parameter choices made one at a time.  See text for
  discussion.}
\label{Fig:extendedh2o}
\end{figure*}

It is also possible and probably more physically realistic to model
the deep cloud as vertically extended instead of as a compact layer.
Such a model, with a uniform optical depth per bar from the 10 bar
level to the bottom of the main (haze2) cloud, can indeed fit the
observations just as well as the prior compact model for the deep
cloud, as long as the cloud density is sufficient to block out thermal
radiation, which requires about 10-20 optical depths per bar. We also
found that larger particles in the \nhtx-containing layer provided a
somewhat better match to the spectral features near 3\mumx, although
maintaining a good overall fit to the spectrum in this case required
adding some absorption at long wavelengths.  To accomplish that, we
made the previously conservative substance absorbing between 4.3 and
5.3 \mum by linearly increasing its imaginary index from zero to 0.03
over that region.  We also decreased the particle radius in the upper
sublayer from 1.5 \mum to 1.0 \mum to further reduce its reflectivity
at longer wavelengths.  The model parameters are provided in
Table\ \ref{Tbl:extendedh2o}.  Some of these were fixed manually,
while others with uncertainty estimates were adjusted to minimize
\chisq within the assumed constraints. This revised model is compared
with observations in Fig.\ \ref{Fig:extendedh2o} and also with the
best linear combination heterogeneous model in Fig.\ \ref{Fig:pure}
(red curve).  While this provides better fit to the spectrum near 3
\mum than the other horizontally homogeneous models, it is still does
not quite match the quality of the best heterogeneous model.  This
probably means that there is more physics needed in the homogeneous
model to capture more complexity in composition and vertical
structure.

\begin{table}\centering
\caption{Fit parameters for storm head Model 3, which includes a vertically extended water cloud.}
\vspace{0.15in}
\begin{tabular}{c c c c c }
\hline\\[-0.1in]
Particle     & \multicolumn{2}{c}{Layer boundaries} & Particle & Optical  \\
 Composition & Ptop, mb   & Pbot, mb   & radius      & depth   \\
\hline\\[-0.1in]
  n=1.4+0i        & 140$\pm$79 & 189$\pm$25 & 1.0 \mum & 1.46$\pm$0.06  \\[0.1in]
\parbox{0.75in}{\nht shell on\newline \hto (core)}     & 189$\pm$25 & 544$\pm$135 & 
         \parbox{.5in}{2.25 \mum \newline (0.84 \mumx)} & 7.1$\pm$1.3 \\[0.15in]
  \hto      &  544$\pm$135  & 10 bars & 2.0 \mum & 20/bar \\[0.1in]
\hline\\[-0.1in]
\end{tabular}
\parbox{3.25in}{Stratospheric haze and \pht scale height as in Table\ \ref{Tbl:waterparams}}.
\label{Tbl:extendedh2o}
\end{table}

\begin{figure*}[!htb]\centering
\includegraphics[width=6.2in]{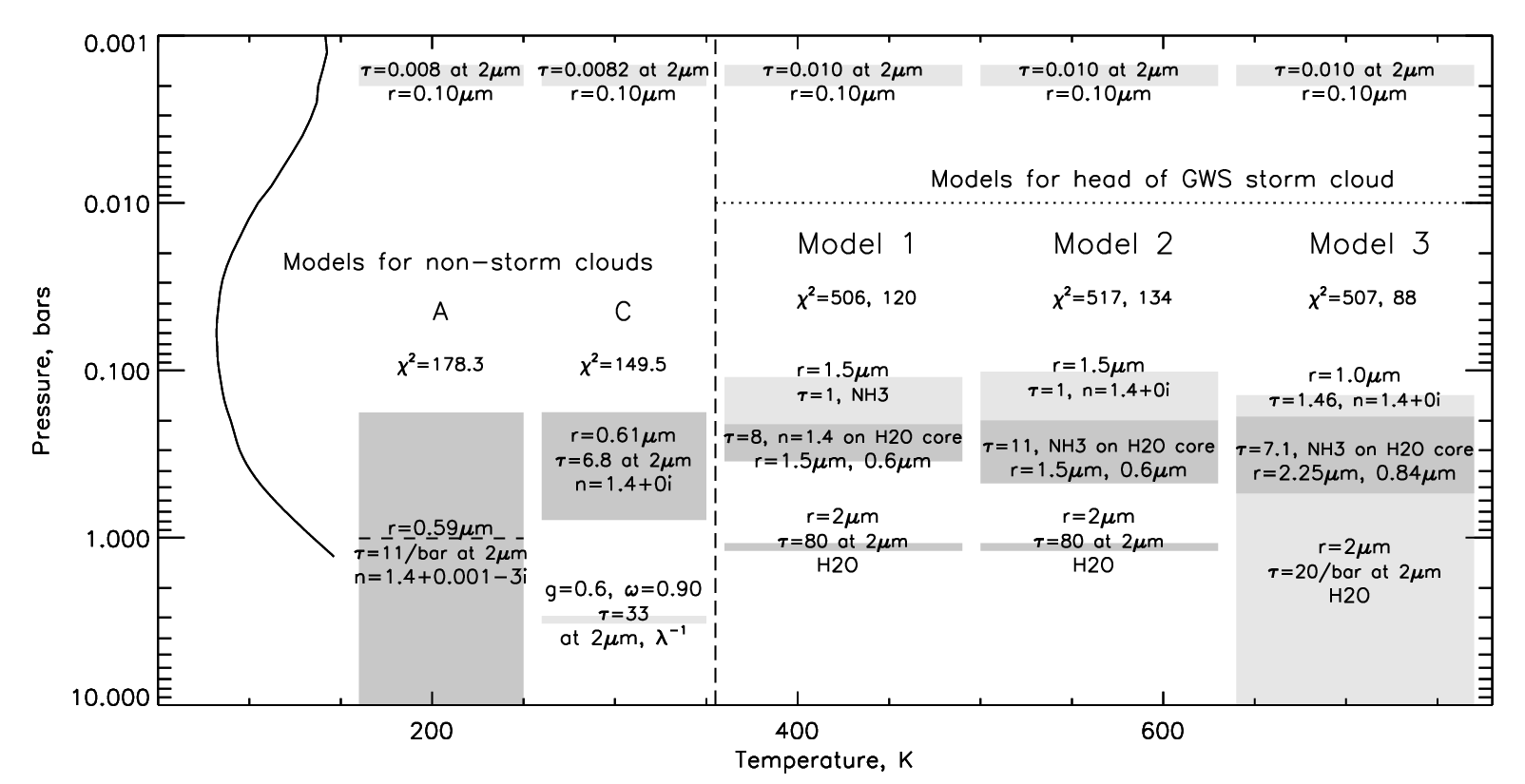}
\caption{Comparison of non-storm cloud structure models A and C (from
  Fig.\ \ref{Fig:cartoon1}) with Models 1-3 for the head of the storm.  The
  two \chisq values given for Models 1-3 are for the 1.268-5.15 \mum
  region (first) and the 2.65-3.2 \mum region (second), with numbers of
comparison points being 251 and 51 respectively.}
\label{Fig:structurecomp}
\end{figure*}

\cite{Sanz-Requena2012} analyzed ground-based visual imaging observations
covering UV and a range of methane bands for the purpose of characterizing
the storm's cloud structure.  Their observations were modeled with a 3-layer
structure consisting of a stratospheric haze, a tropospheric haze, and a bottom
cloud.  They found that the tropospheric haze optical depth was lower in
the storm cloud than it was in the non-storm region (15$\pm$5 vs 25$\pm$5),
although the opacity was concentrated at higher altitudes as they
inferred a cloud boundary change to 100 - 300 mb (storm) from 100 to $>$1000 mb (no storm).
This concentration of optical depth at low pressures is at least roughly consistent
with our inferences from near-IR observations, and also with our finding of a more extended
main layer in the non-storm regions.  However, as shown in Fig.\ \ref{Fig:structurecomp},
we find that the storm region is better fit with a two-component main layer, with most 
of the optical depth in the lower sub-layer, and with the upper sub layer extending
to somewhat lower pressures than found in the non-storm regions.

\section{Summary and conclusions}

From an analysis of VIMS 
spectra of the giant Saturn storm of 2010-2011 and its surroundings, we have
reached the following conclusions.

\begin{enumerate}

\item VIMS spectra outside the storm region exhibit a high degree of
  uniformity in the fractional depth of a 2.95-3.0 \mum absorption feature
  across the Saturn disk, with essentially no dependence on latitude
  or view angle, implying that the feature is primarily due to a
  responsivity error in the vicinity of a 2.98-\mum joint between order-sorting
  filters.  It also has a slight variation with the x coordinate of
  the VIMS images, but essentially no variation with the y coordinate.
  We developed an accurate correction algorithm that removes this feature,
  allowing us to detect true absorption by \nht in the vicinity of
  storm clouds.

\item Most near-IR spectra of Saturn outside the storm regions can be
  very well fit with relatively simple models consisting of (1) a deep
  and often optically thick cloud layer that serves to block thermal
  radiation in most places, but is thin enough, or deep enough, in
  some regions to allow 5-\mum detection of atmospheric composition and
  horizontal structure at the several bar level, (2) a middle ``main''
  cloud layer extending from a top pressure of 111-178 mb to a bottom
  pressure of 577-844 mb, with optical depths of 3.5-8.6 at 2 \mumx,
  and composed of particles that are relatively conservative
  scatterers, with an acceptable refractive index of n=1.4+0i, and (3)
  a stratospheric haze of low optical depth, but which cannot be
  accurately constrained by near-IR VIMS observations because they are
  too uncertain (and generally too large) at very low I/F values (near
  the 1 DN level).

\item Outside the storm-affected regions, there is no evidence for
  particulate absorption at 3 \mumx, further confirming the peculiar
  result that at least the upper several optical depths of Saturn's
  main visible cloud layer is not made of \nhtx, \nhfshx, or \nthfx.
  We cannot currently rule out \pthfx, but its presence seems likely
  to be testable in the future from measurements near 4.3 \mum once
  optical constants for solid diphosphine are determined.

\item Inside the storm head we found evidence for a prominent 3-\mum
  absorber in Saturn's storm clouds but little indication of an
  absorber active near 2 \mumx.  A weak 2-\mum feature is not too
  surprising if the \nht ice is confined to a cloud of relatively
  small optical depth overlying a relatively bright cloud that fills
  in the lack of reflectivity in the absorbing regions of the \nht ice
  spectrum, or if the ammonia-containing cloud layer is underneath a
  conservative scattering layer that serves the same purpose.  An
  uncovered optically thick layer of pure \nht ice particles would
  produce a noticeable feature at 2 \mum as well as a very prominent
  feature 2.96 \mumx, and thus a thick cloud of pure \nht ice
  particles is not plausible as the dominant upper tropospheric cloud
  in the most absorbing storm regions.

\item We considered four plausible candidate materials that might
  contribute significant absorption in the 3-\mum region: \nhtx, \nthfx,
  \nhfshx, and H$_2$O. Fits to the observed spectrum outside of the
  2.8-3.1 \mum region of the spectrum were obtained for a model in
  which each material was taken to be the only component of the main
  cloud layer.  We then compared the features inside 2.8-3.1 \mum part
  of the spectrum as well as outside with the VIMS measured spectrum.
  The only substance that came close to fitting the observed spectrum
  was \nhtx, but it had a very sharp absorption feature at 2.96 \mum
  that was not present in the observed spectrum.

\item We also considered linear combinations of the spectra
obtained from the pure composition fits, as might be obtained
from a heterogeneous horizontal mix of different particle types.
In this case we found that the 2.96-\mum feature could be reduced
in amplitude by adding contributions from other materials.  The
best fits were obtained with a mix of \nhtx, \nhfshx, and H$_2$O, using
areal fractions of 0.55, 0.23, and 0.22 respectively. 

\item The best fits to VIMS spectra using a horizontally homogeneous
  vertical structure model contained a deep water cloud, perhaps
  beginning at 10 bars, but certainly containing significant opacity
  above the 1-2 bar range, a main cloud layer consisting of two
  sub-layers, one of low optical depth ($\tau \approx 1$) above a
  lower sub layer of significant optical depth ($\tau \approx 9$) and
  an optically thin stratospheric haze that we could not reliably
  constrain because of VIMS limitations at very low signal levels.  We
  initially found two options for the main layer produced good fits:
  (1) particles of \nht ice in the top layer and particles of n=1.4
  material (unknown composition also present almost everywhere else on
  Saturn) coating a core of water ice in the lower sublayer, or (2)
  particles of n=1.4 in the top sublayer and \nht in the lower
  sublayer as a coating on \hto cores, with a core radius that is 40\%
  of the total radius (1.5 \mumx), and a core volume of 6\%.  A
  slightly better fit was obtained with option (2) using smaller
  particles with some long-wave absorption in the upper main cloud
  layer and somewhat larger \nhtx-coated water ice particles in the
  lower main sub layer.

\item Although \nhfsh was the best third component of our linear
  combination spectra that represent horizontally heterogeneous
  mixtures, we did not find a useful role for it in our horizontally
  homogeneous composite models.  It might be present in small amounts
  in several layers, but when it becomes a significant fraction of the
  material in the particles, it presents spectral problems that are
  much easier to solve with other materials.  However, it is possible
  that \nhfsh might prove useful as part of a three-component particle
  (core with two coatings), which we do not have the tools to model.

\item Both heterogeneous and homogeneous models of the VIMS storm spectra
strongly favor contributions by water ice, which is the first
strong spectroscopic evidence for water ice in Saturn's atmosphere, found near
the level of Saturn's visible cloud deck where it could only be
delivered by powerful convection originating from $\sim$200 km deeper
in the atmosphere, as expected from numerical dynamical models.

\item The increased scale height we found for \pht in the storm head
relative to the surrounding regions (0.5 compared to 0.375 times the
pressure scale height) is consistent with increased
upward convective transport inside the storm.

\item 
  We find strong evidence for the presence of \nht ice at the top of the
  clouds, but at least some (and perhaps much) of the cloud seems to
  be of the same relatively conservative (at near-IR wavelengths)
  material that forms the main component of clouds over the non-storm
  regions.

\end{enumerate}

\section*{Acknowledgments.} \addcontentsline{toc}{section}{Acknowledgments}

Support for this work was provided by NASA through its Outer Planets
Research Program under grant NNX11AM58G. We thank two anonymous reviewers
for useful suggestions for improving the manuscript.


\begin{thebibliography}{58}
\expandafter\ifx\csname natexlab\endcsname\relax\def\natexlab#1{#1}\fi
\expandafter\ifx\csname url\endcsname\relax
  \def\url#1{\texttt{#1}}\fi
\expandafter\ifx\csname urlprefix\endcsname\relax\def\urlprefix{URL }\fi

\bibitem[{{Acton}(1996)}]{Acton1996}
{Acton}, C.~H., 1996. {Ancillary data services of NASA's Navigation and
  Ancillary Information Facility}. Planet. and Space Sci. 44, 65--70.

\bibitem[{{Baines} et~al.(2009){Baines}, {Delitsky}, {Momary}, {Brown},
  {Buratti}, {Clark}, and {Nicholson}}]{Baines2009stormclouds}
{Baines}, K.~H., {Delitsky}, M.~L., {Momary}, T.~W., {Brown}, R.~H., {Buratti},
  B.~J., {Clark}, R.~N., {Nicholson}, P.~D., 2009. {Storm clouds on Saturn:
  Lightning-induced chemistry and associated materials consistent with
  Cassini/VIMS spectra}. \planss 57, 1650--1658.

\bibitem[{{Baines} et~al.(2011){Baines}, {Momary}, {Fletcher}, {Showman},
  {Brown}, {Buratti}, {Clark}, {Nicholson}, {Go}, and
  {Wesley}}]{Baines2011epsc}
{Baines}, K.~H., {Momary}, T., {Fletcher}, L., {Showman}, A., {Brown}, R.,
  {Buratti}, B., {Clark}, R., {Nicholson}, P., {Go}, C., {Wesley}, A., 2011.
  {Saturn's Enigmatic ''String of Pearls'' and Northern Storm of 2010-2011:
  Manifestations of a Common Dynamical Mechanism?} In: EPSC-DPS Joint Meeting
  2011. p. 1658.

\bibitem[{{Birnbaum} et~al.(1996){Birnbaum}, {Borysow}, and
  {Orton}}]{Birnbaum1996}
{Birnbaum}, G., {Borysow}, A., {Orton}, G.~S., 1996. {Collision-Induced
  Absorption of H$_2$-H$_2$ and H$_2$-He in the Rotational and Fundamental
  Bands for Planetary Applications}. Icarus 123, 4--22.

\bibitem[{{Borysow}(1991)}]{Borysow1991h2h2f}
{Borysow}, A., 1991. {Modeling of collision-induced infrared absorption spectra
  of H$_2$-H$_2$ pairs in the fundamental band at temperatures from 20 to 300
  K}. Icarus 92, 273--279.

\bibitem[{{Borysow}(1992)}]{Borysow1992h2he}
{Borysow}, A., 1992. {New model of collision-induced infrared absorption
  spectra of H$_2$-He pairs in the 2-2.5 micron range at temperatures from 20
  to 300 K - an update}. Icarus 96, 169--175.

\bibitem[{{Borysow}(1993)}]{Borysow1993errat}
{Borysow}, A., 1993. {Erratum}. Icarus 106, 614.

\bibitem[{{Bowles} et~al.(2008){Bowles}, {Calcutt}, {Irwin}, and
  {Temple}}]{Bowles2008}
{Bowles}, N., {Calcutt}, S., {Irwin}, P., {Temple}, J., 2008. {Band parameters
  for self-broadened ammonia gas in the range 0.74 to 5.24 {$\mu$}m to support
  measurements of the atmosphere of the planet Jupiter}. Icarus 196, 612--624.

\bibitem[{{Briggs} and {Sackett}(1989)}]{Briggs1989}
{Briggs}, F.~H., {Sackett}, P.~D., 1989. {Radio observations of Saturn as a
  probe of its atmosphere and cloud structure}. Icarus 80, 77--103.

\bibitem[{{Brown} et~al.(2004){Brown}, {Baines}, {Bellucci}, {Bibring},
  {Buratti}, {Capaccioni}, {Cerroni}, {Clark}, {Coradini}, {Cruikshank},
  {Drossart}, {Formisano}, {Jaumann}, {Langevin}, {Matson}, {McCord},
  {Mennella}, {Miller}, {Nelson}, {Nicholson}, {Sicardy}, and
  {Sotin}}]{Brown2004SSR}
{Brown}, R.~H., {Baines}, K.~H., {Bellucci}, G., {Bibring}, J.-P., {Buratti},
  B.~J., {Capaccioni}, F., {Cerroni}, P., {Clark}, R.~N., {Coradini}, A.,
  {Cruikshank}, D.~P., {Drossart}, P., {Formisano}, V., {Jaumann}, R.,
  {Langevin}, Y., {Matson}, D.~L., {McCord}, T.~B., {Mennella}, V., {Miller},
  E., {Nelson}, R.~M., {Nicholson}, P.~D., {Sicardy}, B., {Sotin}, C., 2004.
  {The Cassini Visual and Infrared Mapping Spectrometer (VIMS) Investigation}.
  Space Science Reviews 115, 111--168.

\bibitem[{{Butler} et~al.(2006){Butler}, {Sagui}, {Kleiner}, and
  {Brown}}]{Butler2006}
{Butler}, R.~A.~H., {Sagui}, L., {Kleiner}, I., {Brown}, L.~R., 2006. {The
  absorption spectrum of phosphine (PH$_{3}$) between 2.8 and 3.7 {$\mu$}m:
  Line positions, intensities, and assignments}. J. of Molec. Spectrosc. 238,
  178--192.

\bibitem[{{Campargue} et~al.(2012){Campargue}, {Leshchishina}, {Wang},
  {Mondelain}, , {Kassi}, and {Nikitin}}]{Campargue2012refine}
{Campargue}, A., {Leshchishina}, O., {Wang}, L., {Mondelain}, D., , {Kassi},
  S., {Nikitin}, A., 2012. {Refinements of the WKMC empirical line lists
  (5852-7919 cm$^{-1}$) for methane between 80 K and 296 K}. \jqsrt 113,
  1855--1873.

\bibitem[{{Clapp} and {Miller}(1996)}]{Clapp1996}
{Clapp}, M.~L., {Miller}, R.~E., 1996. {Complex Refractive Indices of
  Crystalline Hydrazine from Aerosol Extinction Spectra}. Icarus 123, 396--403.

\bibitem[{{Colina} et~al.(1996){Colina}, {Bohlin}, and {Castelli}}]{Colina1996}
{Colina}, L., {Bohlin}, R.~C., {Castelli}, F., 1996. {The 0.12-2.5 micron
  Absolute Flux Distribution of the Sun for Comparison With Solar Analog
  Stars}. \aj 112, 307--315.

\bibitem[{{Conrath} and {Gautier}(2000)}]{Conrath2000}
{Conrath}, B.~J., {Gautier}, D., 2000. {Saturn Helium Abundance: A Reanalysis
  of Voyager Measurements}. Icarus 144, 124--134.

\bibitem[{{Courtin} et~al.(1984){Courtin}, {Gautier}, {Marten}, {B\'ezard}, and
  {Hanel}}]{Courtin1984}
{Courtin}, R., {Gautier}, D., {Marten}, A., {B\'ezard}, B., {Hanel}, R., 1984.
  {The composition of Saturn's atmosphere at northern temperate latitudes from
  Voyager IRIS spectra - NH$_3$, PH$_3$, C$_2$H$_2$, C$_2$H$_6$, CH$_3$D,
  CH$_4$, and the Saturnian D/H isotopic ratio}. \apj 287, 899--916.

\bibitem[{{Cruikshank} et~al.(2010){Cruikshank}, {Meyer}, {Brown}, {Clark},
  {Jaumann}, {Stephan}, {Hibbitts}, {Sandford}, {Mastrapa}, {Filacchione},
  {Ore}, {Nicholson}, {Buratti}, {McCord}, {Nelson}, {Dalton}, {Baines}, and
  {Matson}}]{Cruikshank2010}
{Cruikshank}, D.~P., {Meyer}, A.~W., {Brown}, R.~H., {Clark}, R.~N., {Jaumann},
  R., {Stephan}, K., {Hibbitts}, C.~A., {Sandford}, S.~A., {Mastrapa},
  R.~M.~E., {Filacchione}, G., {Ore}, C.~M.~D., {Nicholson}, P.~D., {Buratti},
  B.~J., {McCord}, T.~B., {Nelson}, R.~M., {Dalton}, J.~B., {Baines}, K.~H.,
  {Matson}, D.~L., 2010. {Carbon dioxide on the satellites of Saturn: Results
  from the Cassini VIMS investigation and revisions to the VIMS wavelength
  scale}. Icarus 206, 561--572.

\bibitem[{{Dyudina} et~al.(2010){Dyudina}, {Ingersoll}, {Ewald}, {Porco},
  {Fischer}, {Kurth}, and {West}}]{Dyudina2010GRL}
{Dyudina}, U.~A., {Ingersoll}, A.~P., {Ewald}, S.~P., {Porco}, C.~C.,
  {Fischer}, G., {Kurth}, W.~S., {West}, R.~A., 2010. {Detection of visible
  lightning on Saturn}. \grl 37, L09205.

\bibitem[{{Encrenaz} et~al.(1999){Encrenaz}, {Drossart}, {Feuchtgruber},
  {Lellouch}, {B{\'e}zard}, {Fouchet}, and {Atreya}}]{Encrenaz1999}
{Encrenaz}, T., {Drossart}, P., {Feuchtgruber}, H., {Lellouch}, E.,
  {B{\'e}zard}, B., {Fouchet}, T., {Atreya}, S.~K., 1999. {The atmospheric
  composition and structure of Jupiter and Saturn from ISO observations: a
  preliminary review}. Plan. \& Sp. Sci. 47, 1225--1242.

\bibitem[{{Evans} and {Stephens}(1991)}]{Evans1991}
{Evans}, K.~F., {Stephens}, G.~L., 1991. {A new polarized atmospheric radiative
  transfer model}. J. Quant. Spectr. and Rad. Trans. 46, 413--423.

\bibitem[{{Fletcher} et~al.(2011{\natexlab{a}}){Fletcher}, {Baines}, {Momary},
  {Showman}, {Irwin}, {Orton}, M., and {Merlit}}]{Fletcher2011vims}
{Fletcher}, L.~N., {Baines}, K.~H., {Momary}, T.~M., {Showman}, A.~S., {Irwin},
  P.~G.~J., {Orton}, G.~S., M., R., {Merlit}, C., 2011{\natexlab{a}}. {Saturn's
  Tropospheric Composition and Clouds from Cassini/VIMS 4.6-5.1 $\mu$m
  Nightside Spectroscopy}. Icarus 214, 510--533.

\bibitem[{{Fletcher} et~al.(2011{\natexlab{b}}){Fletcher}, {Hesman}, {Irwin},
  {Baines}, {Momary}, {Sanchez-Lavega}, {Flasar}, {Read}, {Orton},
  {Simon-Miller}, {Hueso}, {Bjoraker}, {Mamoutkine}, {del Rio-Gaztelurrutia},
  {Gomez}, {Buratti}, {Clark}, {Nicholson}, and {Sotin}}]{Fletcher2011Sci}
{Fletcher}, L.~N., {Hesman}, B.~E., {Irwin}, P.~G.~J., {Baines}, K.~H.,
  {Momary}, T.~W., {Sanchez-Lavega}, A., {Flasar}, F.~M., {Read}, P.~L.,
  {Orton}, G.~S., {Simon-Miller}, A., {Hueso}, R., {Bjoraker}, G.~L.,
  {Mamoutkine}, A., {del Rio-Gaztelurrutia}, T., {Gomez}, J.~M., {Buratti}, B.,
  {Clark}, R.~N., {Nicholson}, P.~D., {Sotin}, C., 2011{\natexlab{b}}. {Thermal
  Structure and Dynamics of Saturn's Northern Springtime Disturbance}. Science
  332, 1413--1417.

\bibitem[{{Fletcher} et~al.(2009{\natexlab{a}}){Fletcher}, {Orton}, {Teanby},
  and {Irwin}}]{Fletcher2009ph3}
{Fletcher}, L.~N., {Orton}, G.~S., {Teanby}, N.~A., {Irwin}, P.~G.~J.,
  2009{\natexlab{a}}. {Phosphine on Jupiter and Saturn from Cassini/CIRS}.
  Icarus 202, 543--564.

\bibitem[{{Fletcher} et~al.(2009{\natexlab{b}}){Fletcher}, {Orton}, {Teanby},
  {Irwin}, and {Bjoraker}}]{Fletcher2009ch4saturn}
{Fletcher}, L.~N., {Orton}, G.~S., {Teanby}, N.~A., {Irwin}, P.~G.~J.,
  {Bjoraker}, G.~L., 2009{\natexlab{b}}. {Methane and its isotopologues on
  Saturn from Cassini/CIRS observations}. Icarus 199, 351--367.

\bibitem[{{Fouchet} et~al.(2009){Fouchet}, {Moses}, and
  {Conrath}}]{Fouchet2009}
{Fouchet}, T., {Moses}, J.~I., {Conrath}, B.~J., 2009. {Saturn: Composition and
  Chemistry}. In: {Dougherty}, M.~K., {Esposito}, L.~W., {Krimigis}, S.~M.
  (Eds.), Saturn from Cassini-Huygens. {Springer, Dordrecht, Heidelberg,
  London, New York}, pp. 83--112.

\bibitem[{{Frankiss}(1968)}]{Frankiss1968}
{Frankiss}, S.~G., 1968. {Vibrational Spectrum and Structure of Solid
  Diphosphine}. Inorg. Chem. 7, 1931--1933.

\bibitem[{{Gautier} et~al.(2006){Gautier}, {Conrath}, {Flasar}, {Achterberg},
  {Schinder}, {Kliore}, {Cassini Cirs}, and {Radio Science
  Teams}}]{Gautier2006cosp}
{Gautier}, D., {Conrath}, B., {Flasar}, M., {Achterberg}, R., {Schinder}, P.,
  {Kliore}, A., {Cassini Cirs}, {Radio Science Teams}, 2006. {The helium to
  hydrogen ratio in Saturn's atmosphere from Cassini CIRS and radio science
  measurement}. In: 36th COSPAR Scientific Assembly. Vol.~36 of COSPAR Meeting.
  p. 867.

\bibitem[{{Howett} et~al.(2007){Howett}, {Carlson}, {Irwin}, and
  {Calcutt}}]{Howett2007}
{Howett}, C.~J.~A., {Carlson}, R.~W., {Irwin}, P.~G.~J., {Calcutt}, S.~B.,
  2007. {Optical constants of ammonium hydrosulfide ice and ammonia ice}.
  Journal of the Optical Society of America B Optical Physics 24, 126--136.

\bibitem[{{Hueso} and {S{\'a}nchez-Lavega}(2004)}]{Hueso2004Icar}
{Hueso}, R., {S{\'a}nchez-Lavega}, A., 2004. {A three-dimensional model of
  moist convection for the giant planets II: Saturn's water and ammonia moist
  convective storms}. Icarus 172, 255--271.

\bibitem[{{Irwin} et~al.(2006){Irwin}, {Sromovsky}, {Strong}, {Sihra},
  {Bowles}, {Calcutt}, and {Remedios}}]{Irwin2006ch42e}
{Irwin}, P.~G.~J., {Sromovsky}, L.~A., {Strong}, E.~K., {Sihra}, K., {Bowles},
  N., {Calcutt}, S.~B., {Remedios}, J.~J., 2006. {Improved near-infrared
  methane band models and k-distribution parameters from 2000 to 9500 cm$^{-1}$
  and implications for interpretation of outer planet spectra}. Icarus 181,
  309--319.

\bibitem[{{Karkoschka} and {Tomasko}(2005)}]{Kark2005Icar}
{Karkoschka}, E., {Tomasko}, M., 2005. {Saturn's vertical and latitudinal cloud
  structure 1991-2004 from HST imaging in 30 filters}. Icarus 179, 195--221.

\bibitem[{{Lacis} and {Oinas}(1991)}]{Lacis1991}
{Lacis}, A.~A., {Oinas}, V., 1991. {A description of the correlated-k
  distribution method for modelling nongray gaseous absorption, thermal
  emission, and multiple scattering in vertically inhomogeneous atmospheres}.
  \jgr 96, 9027--9064.

\bibitem[{{Lindal} et~al.(1985){Lindal}, {Sweetnam}, and
  {Eshleman}}]{Lindal1985}
{Lindal}, G.~F., {Sweetnam}, D.~N., {Eshleman}, V.~R., 1985. {The atmosphere of
  Saturn - an analysis of the Voyager radio occultation measurements}. \aj 90,
  1136--1146.

\bibitem[{{Martonchik} et~al.(1984){Martonchik}, {Orton}, and
  {Appleby}}]{Martonchik1984}
{Martonchik}, J.~V., {Orton}, G.~S., {Appleby}, J.~F., 1984. {Optical
  properties of NH$_3$ ice from the far infrared to the near ultraviolet}.
  Appl. Optics 23, 541--547.

\bibitem[{{McCord} et~al.(2004){McCord}, {Coradini}, {Hibbitts}, {Capaccioni},
  {Hansen}, {Filacchione}, {Clark}, {Cerroni}, {Brown}, {Baines}, {Bellucci},
  {Bibring}, {Buratti}, {Bussoletti}, {Combes}, {Cruikshank}, {Drossart},
  {Formisano}, {Jaumann}, {Langevin}, {Matson}, {Nelson}, {Nicholson},
  {Sicardy}, and {Sotin}}]{McCord2004}
{McCord}, T.~B., {Coradini}, A., {Hibbitts}, C.~A., {Capaccioni}, F., {Hansen},
  G.~B., {Filacchione}, G., {Clark}, R.~N., {Cerroni}, P., {Brown}, R.~H.,
  {Baines}, K.~H., {Bellucci}, G., {Bibring}, J.-P., {Buratti}, B.~J.,
  {Bussoletti}, E., {Combes}, M., {Cruikshank}, D.~P., {Drossart}, P.,
  {Formisano}, V., {Jaumann}, R., {Langevin}, Y., {Matson}, D.~L., {Nelson},
  R.~M., {Nicholson}, P.~D., {Sicardy}, B., {Sotin}, C., 2004. {Cassini VIMS
  observations of the Galilean satellites including the VIMS calibration
  procedure}. Icarus 172, 104--126.

\bibitem[{{Miller} et~al.(1996){Miller}, {Klein}, {Juergens}, {Mehaffey},
  {Oseas}, {Garcia}, {Giandomenico}, {Irigoyen}, {Hickok}, {Rosing}, {Sobel},
  {Bruce}, {Flamini}, {Devidi}, {Reininger}, {Dami}, {Soufflot}, {Langevin},
  and {Huntzinger}}]{Miller1996SPIE}
{Miller}, E.~A., {Klein}, G., {Juergens}, D.~W., {Mehaffey}, K., {Oseas},
  J.~M., {Garcia}, R.~A., {Giandomenico}, A., {Irigoyen}, R.~E., {Hickok}, R.,
  {Rosing}, D., {Sobel}, H.~R., {Bruce}, C.~F., {Flamini}, E., {Devidi}, R.,
  {Reininger}, F.~M., {Dami}, M., {Soufflot}, A., {Langevin}, Y., {Huntzinger},
  G., 1996. {The Visual and Infrared Mapping Spectrometer for Cassini}. In:
  {Horn}, L. (Ed.), Society of Photo-Optical Instrumentation Engineers (SPIE)
  Conference Series. Vol. 2803 of Society of Photo-Optical Instrumentation
  Engineers (SPIE) Conference Series. pp. 206--220.

\bibitem[{{Mu{\~n}oz} et~al.(2004){Mu{\~n}oz}, {Moreno}, {Molina}, {Grodent},
  {G{\'e}rard}, and {Dols}}]{Munoz2004}
{Mu{\~n}oz}, O., {Moreno}, F., {Molina}, A., {Grodent}, D., {G{\'e}rard},
  J.~C., {Dols}, V., 2004. {Study of the vertical structure of Saturn's
  atmosphere using HST/WFPC2 images}. Icarus 169, 413--428.

\bibitem[{{Nixon}(1956)}]{Nixon1956}
{Nixon}, E.~R., 1956. {The Infrared Spectrum of Biphosphine}. J. Phys. Chem.
  60, 1054--1059.

\bibitem[{{P{\'e}rez-Hoyos} et~al.(2005){P{\'e}rez-Hoyos},
  {S{\'a}nchez-Lavega}, {French}, and {Rojas}}]{Perez-Hoyos2005}
{P{\'e}rez-Hoyos}, S., {S{\'a}nchez-Lavega}, A., {French}, R.~G., {Rojas},
  J.~F., 2005. {Saturn's cloud structure and temporal evolution from ten years
  of Hubble Space Telescope images (1994 2003)}. Icarus 176, 155--174.

\bibitem[{{Press} et~al.(1992){Press}, {Teukolsky}, {Vetterling}, and
  {Flannery}}]{Press1992}
{Press}, W.~H., {Teukolsky}, S.~A., {Vetterling}, W.~T., {Flannery}, B.~P.,
  1992. {Numerical recipes in FORTRAN. The art of scientific computing, 2nd
  ed.} Cambridge: University Press.

\bibitem[{{Prinn} et~al.(1984){Prinn}, {Larson}, {Caldwell}, and
  {Gautier}}]{Prinn1984}
{Prinn}, R.~G., {Larson}, H.~P., {Caldwell}, J.~J., {Gautier}, D., 1984.
  {Composition and chemistry of Saturn's atmosphere}. In: {Gehrels, T.~\&
  Matthews, M.~S.} (Ed.), Saturn. Univ. of Arizona Press, Tucson, pp. 88--149.

\bibitem[{{Rice}(1995)}]{Rice1995}
{Rice}, J.~A., 1995. {Mathematical Statistics and Data Analysis}, 2nd Edition.
  Duxbury Press, Belmont, California.

\bibitem[{{Rothman} et~al.(2009){Rothman}, {Gordon}, {Barbe}, {Benner},
  {Bernath}, {Birk}, {Boudon}, {Brown}, {Campargue}, {Champion}, {Chance},
  {Coudert}, {Dana}, {Devi}, {Fally}, {Flaud}, {Gamache}, {Goldman},
  {Jacquemart}, {Kleiner}, {Lacome}, {Lafferty}, {Mandin}, {Massie},
  {Mikhailenko}, {Miller}, {Moazzen-Ahmadi}, {Naumenko}, {Nikitin}, {Orphal},
  {Perevalov}, {Perrin}, {Predoi-Cross}, {Rinsland}, {Rotger}, {{\v S}ime{\v
  c}kov{\'a}}, {Smith}, {Sung}, {Tashkun}, {Tennyson}, {Toth}, {Vandaele}, and
  {Vander Auwera}}]{Rothman2009}
{Rothman}, L.~S., {Gordon}, I.~E., {Barbe}, A., {Benner}, D.~C., {Bernath},
  P.~F., {Birk}, M., {Boudon}, V., {Brown}, L.~R., {Campargue}, A., {Champion},
  J., {Chance}, K., {Coudert}, L.~H., {Dana}, V., {Devi}, V.~M., {Fally}, S.,
  {Flaud}, J., {Gamache}, R.~R., {Goldman}, A., {Jacquemart}, D., {Kleiner},
  I., {Lacome}, N., {Lafferty}, W.~J., {Mandin}, J., {Massie}, S.~T.,
  {Mikhailenko}, S.~N., {Miller}, C.~E., {Moazzen-Ahmadi}, N., {Naumenko},
  O.~V., {Nikitin}, A.~V., {Orphal}, J., {Perevalov}, V.~I., {Perrin}, A.,
  {Predoi-Cross}, A., {Rinsland}, C.~P., {Rotger}, M., {{\v S}ime{\v
  c}kov{\'a}}, M., {Smith}, M.~A.~H., {Sung}, K., {Tashkun}, S.~A., {Tennyson},
  J., {Toth}, R.~A., {Vandaele}, A.~C., {Vander Auwera}, J., 2009. {The HITRAN
  2008 molecular spectroscopic database}. \jqsrt 110, 533--572.

\bibitem[{{Roux} et~al.(1979){Roux}, {Wood}, and {Smith}}]{Roux1979}
{Roux}, J.~A., {Wood}, B.~E., {Smith}, A.~M., 1979. {Optical properties of thin
  H$_2$0, NH$_3$, and CO$_2$ cryofilms. AEDC-TR-79-57}. Arnold Engineering
  Development Center, Tennessee.

\bibitem[{{S\'anchez-Lavega} et~al.(1991){S\'anchez-Lavega}, {Colas},
  {Lecacheux}, {Laques}, {Parker}, and {Miyazaki}}]{Sanchez-Lavega1991}
{S\'anchez-Lavega}, A., {Colas}, F., {Lecacheux}, J., {Laques}, P., {Parker},
  D., {Miyazaki}, I., 1991. {The Great White Spot and disturbances in Saturn's
  equatorial atmosphere during 1990}. Nature 353, 397--401.

\bibitem[{{S{\'a}nchez-Lavega} et~al.(2011){S{\'a}nchez-Lavega}, {del
  R{\'{\i}}o-Gaztelurrutia}, {Hueso}, {G{\'o}mez-Forrellad}, {Sanz-Requena},
  {Legarreta}, {Garc{\'{\i}}a-Melendo}, {Colas}, {Lecacheux}, {Fletcher},
  {Barrado y Navascu{\'e}s}, {Parker}, {International Outer Planet Watch Team},
  {Akutsu}, {Barry}, {Beltran}, {Buda}, {Combs}, {Carvalho}, {Casquinha},
  {Delcroix}, {Ghomizadeh}, {Go}, {Hotershall}, {Ikemura}, {Jolly}, {Kazemoto},
  {Kumamori}, {Lecompte}, {Maxson}, {Melillo}, {Milika}, {Morales}, {Peach},
  {Phillips}, {Poupeau}, {Sussenbach}, {Walker}, {Walker}, {Tranter}, {Wesley},
  {Wilson}, and {Yunoki}}]{Sanchez-Lavega2011Nature}
{S{\'a}nchez-Lavega}, A., {del R{\'{\i}}o-Gaztelurrutia}, T., {Hueso}, R.,
  {G{\'o}mez-Forrellad}, J.~M., {Sanz-Requena}, J.~F., {Legarreta}, J.,
  {Garc{\'{\i}}a-Melendo}, E., {Colas}, F., {Lecacheux}, J., {Fletcher}, L.~N.,
  {Barrado y Navascu{\'e}s}, D., {Parker}, D., {International Outer Planet
  Watch Team}, {Akutsu}, T., {Barry}, T., {Beltran}, J., {Buda}, S., {Combs},
  B., {Carvalho}, F., {Casquinha}, P., {Delcroix}, M., {Ghomizadeh}, S., {Go},
  C., {Hotershall}, J., {Ikemura}, T., {Jolly}, G., {Kazemoto}, A., {Kumamori},
  T., {Lecompte}, M., {Maxson}, P., {Melillo}, F.~J., {Milika}, D.~P.,
  {Morales}, E., {Peach}, D., {Phillips}, J., {Poupeau}, J.~J., {Sussenbach},
  J., {Walker}, G., {Walker}, S., {Tranter}, T., {Wesley}, A., {Wilson}, T.,
  {Yunoki}, K., 2011. {Deep winds beneath Saturn's upper clouds from a seasonal
  long-lived planetary-scale storm}. Nature 475, 71--74.

\bibitem[{{Sanz-Requena} et~al.(2012){Sanz-Requena}, {P{\'e}rez-Hoyos},
  {S{\'a}nchez-Lavega}, {del R{\'{\i}}o-Gaztelurrutia},
  {Barrado-Navascu{\'e}s}, {Colas}, {Lecacheux}, and
  {Parker}}]{Sanz-Requena2012}
{Sanz-Requena}, J.~F., {P{\'e}rez-Hoyos}, S., {S{\'a}nchez-Lavega}, A., {del
  R{\'{\i}}o-Gaztelurrutia}, T., {Barrado-Navascu{\'e}s}, D., {Colas}, F.,
  {Lecacheux}, J., {Parker}, D., 2012. {Cloud structure of Saturn's 2010 storm
  from ground-based visual imaging}. Icarus 219, 142--149.

\bibitem[{{Sill} et~al.(1980){Sill}, {Fink}, and {Ferraro}}]{Sill1980}
{Sill}, G., {Fink}, U., {Ferraro}, J.~R., 1980. {Absorption coefficients of
  solid NH3 from 50 to 7000 per cm}. Opt. Soc. of Am. J. A 70, 724--739.

\bibitem[{{Sromovsky}(2005{\natexlab{a}})}]{Sro2005raman}
{Sromovsky}, L.~A., 2005{\natexlab{a}}. {Accurate and approximate calculations
  of Raman scattering in the atmosphere of Neptune}. Icarus 173, 254--283.

\bibitem[{{Sromovsky}(2005{\natexlab{b}})}]{Sro2005pol}
{Sromovsky}, L.~A., 2005{\natexlab{b}}. {Effects of Rayleigh-scattering
  polarization on reflected intensity: a fast and accurate approximation method
  for atmospheres with aerosols}. Icarus 173, 284--294.

\bibitem[{{Sromovsky} and {Fry}(2010{\natexlab{a}})}]{Sro2010iso}
{Sromovsky}, L.~A., {Fry}, P.~M., 2010{\natexlab{a}}. {The source of 3-{$\mu$}m
  absorption in Jupiter's clouds: Reanalysis of ISO observations using new
  NH$_{3}$ absorption models}. Icarus 210, 211--229.

\bibitem[{{Sromovsky} and {Fry}(2010{\natexlab{b}})}]{Sro2010vims}
{Sromovsky}, L.~A., {Fry}, P.~M., 2010{\natexlab{b}}. {The source of widespread
  3-{$\mu$}m absorption in Jupiter's clouds: Constraints from 2000 Cassini VIMS
  observations}. Icarus 210, 230--257.

\bibitem[{{Sromovsky} et~al.(2012){Sromovsky}, {Fry}, {Boudon}, {Campargue},
  and {Nikitin}}]{Sro2012LBL}
{Sromovsky}, L.~A., {Fry}, P.~M., {Boudon}, V., {Campargue}, A., {Nikitin}, A.,
  2012. {Comparison of line-by-line and band models of near-IR methane
  absorption applied to outer planet atmospheres}. Icarus 218, 1--23.

\bibitem[{{Tarrago}(1996)}]{Tarrago1996}
{Tarrago}, G., 1996. {Ground State Rotational Energies of Arsine}. J. of Mol.
  Spectr. 178, 10--21.

\bibitem[{{Toon} and {Ackerman}(1981)}]{Toon1981}
{Toon}, O.~B., {Ackerman}, T.~P., 1981. {Algorithms for the calculation of
  scattering by stratified spheres}. \ao 20, 3657--3660.

\bibitem[{{Warren}(1984)}]{Warren1984}
{Warren}, S.~G., 1984. {Optical constants of ice from the ultraviolet to the
  microwave}. Appl. Optics 23, 1206--1225.

\bibitem[{{West} et~al.(2009){West}, {Baines}, {Karkoschka}, and
  {S{\'a}nchez-Lavega}}]{West2009satbook}
{West}, R.~A., {Baines}, K.~H., {Karkoschka}, E., {S{\'a}nchez-Lavega}, A.,
  2009. {Clouds and Aerosols in Saturn's Atmosphere}. In: Dougherty, M.~K.,
  Esposito, L.~W., Krimigis, S.~M. (Eds.), {Saturn from Cassini-Huygens}.
  {Springer}, pp. 161--179.

\bibitem[{{Zheng} and {Borysow}(1995)}]{Zheng1995h2h2o1}
{Zheng}, C., {Borysow}, A., 1995. {Modeling of collision-induced infrared
  absorption spectra of H$_2$ pairs in the first overtone band at temperatures
  from 20 to 500 K}. Icarus 113, 84--90.

\end{thebibliography}

\end{document}